\documentclass[prl,twocolumn,superscriptaddress,showpacs]{revtex4}
\usepackage{graphicx}
\usepackage{amssymb,amsmath}
\usepackage[T1]{fontenc}
\begin{document}

\title{Stripe phases - possible ground state of 
       the high-$T_c$ superconductors}

\author{Marcin Raczkowski}
\affiliation{Marian Smoluchowski Institute of Physics, Jagellonian
             University, Reymonta 4, PL-30059 Krak\'ow, Poland}
\affiliation{Laboratoire CRISMAT, UMR CNRS--ENSICAEN(ISMRA) 6508, 
             6, Bld. du Mar\'echal Juin, F-14050 Caen, France}

\author{Andrzej M. Ole\'s}
\affiliation{Marian Smoluchowski Institute of Physics, Jagellonian
             University, Reymonta 4, PL-30059 Krak\'ow, Poland}

\author{Raymond Fr\'esard}
\affiliation{Laboratoire CRISMAT, UMR CNRS--ENSICAEN(ISMRA) 6508, 
             6, Bld. du Mar\'echal Juin, F-14050 Caen, France}

\date{\today}

\begin{abstract}
Based on the mean-field method applied either to the extended single-band 
Hubbard model or to the single-band Peierls-Hubbard Hamiltonian we
study the stability of both site-centered and bond-centered charge domain 
walls. The difference in energy between these phases is found to be 
small. Therefore, moderate perturbations to the pure Hubbard model, such
as next nearest hopping, lattice anisotropy, or coupling to the lattice, 
induce phase transitions, shown in the corresponding phase diagrams. 
In addition, we determine for stable phases charge and magnetization 
densities, double occupancy, kinetic and magnetic energies, and 
investigate the role of a finite electron-lattice coupling. We also 
review experimental signatures of stripes in the superconducting copper 
oxides.
\end{abstract}

\maketitle

\section{\label{sec:1} Introduction}

Since the discovery of high-temperature superconductivity by Bednorz and 
M\"uller \cite{Bed86}, the unusual physical properties of the copper 
oxides have stimulated theorists and have led to the appearance of many 
new ideas \cite{Car02}. One of the especially appealing new pictures 
that has emerged is the instability towards a novel type of coexisting 
incommensurate (IC) charge and magnetic order, i.e., stripe phase. As a
rare event in the theory of high temperature superconductivity, the 
theory preceeded here the experiment and the existence of stripe 
phases was predicted on the basis of Hartree-Fock (HF) calculations in 
the two-band model for CuO$_2$ planes of layered La$_{2-x}$Sr$_x$CuO$_4$ 
(LSCO) \cite{Zaa89}, before their experimental confirmation. This 
instability persists as well in the effective single-band Hubbard model 
\cite{Poi89,Sch89,Kat90,Inu91}. All
these calculations yielded solutions with a phase separation manifested 
in formation of nonmagnetic lines of holes, one-dimensional (1D) domain 
walls or stripes, which separate antiferromagnetic (AF) domains of 
opposite phases. Such states result from the competition between the 
superexchange interaction, which stabilize the AF long-range order in 
the parent Mott insulator, and the kinetic energy of doped holes. Indeed, 
the magnetic energy is gained when electrons occupy the neighboring sites 
and their spins order as in the N\'eel state, whereas the kinetic energy 
is gained when the holes can move and the AF order is locally suppressed 
along a domain wall (DW). Thus, a stripe phase provides the best 
compromise between the superexchange promoting the AF order and the 
kinetic energy of doped holes. 

However, the debate on the microscopic origin of the stripe instability 
is far from closed. Two main scenarios, based on a Ginzburg-Landau free 
energy, for the driving mechanism of the stripe phase have been discussed 
\cite{Zac98, Kiv96}. In the first one, stripes are charge-density waves 
with large periodicity arising from the Fermi surface (FS) instability 
with the transition being spin driven \cite{Zaa89}. A general feature of  
such an instability is a gap/pseudogap which opens up precisely 
on the FS. Hence, the spacing between DWs is equal to $1/x$, 
with $x$ denoting doping level so as to maintain a gap/pseudogap on the 
FS. In this scenario spin and charge order occur at the same 
temperature or charge stripe order sets in only after spin order 
has developed. 

An alternative scenario comes from the Coulomb-frustrated phase
separation suggesting that stripe formation is charge driven. Indeed, 
using the Ising model, it has been shown that the competition between 
long range Coulomb interactions and short range attraction 
between holes leads to formation of stripes \cite{Low94}.   
In this case Ginzburg-Landau considerations lead to an onset of charge 
order prior to spin order as the temperature is lowered.
However, the above analysis does not take into account spin fluctuations 
which might be crucial for the nature of the phase transition by 
precluding the spins from ordering at the charge-order temperature 
\cite{Dui98}. Moreover, the conjecture that long range Coulomb forces 
are required to stabilize stripe phases has been challenged by the studies 
of the $t$-$J$ model, in which the DW structures were obtained without 
such interactions \cite{Whi98}.   

In order to investigate the influence of strong electron correlations
due to large on-site Coulomb repulsion $U$ at Cu ions, several methods 
have been employed to study the stripe phases which go beyond the HF 
approximation, such as: 
Density Matrix Renormalization Group (DMRG) \cite{Whi98,Whi99}, 
Slave-Boson Approximation (SBA) \cite{Sei98,Sei98V,Sei04}, 
variational local ansatz approximation \cite{Gor99}, 
Exact Diagonalization (ED) of finite clusters \cite{Toh99}, 
analytical approach based on variational trial wave 
function within the string picture \cite{Wro00}, 
Dynamical Mean Field Theory (DMFT) \cite{Fle00,Fle01},  
Cluster Perturbation Theory (CPT) \cite{Zac00},
and Quantum Monte Carlo (QMC) \cite{Bec01,Rie01}.  
In spite of this huge effort, it remains unclear whether DWs are centered 
on rows of metal atoms, hereafter named site-centered (SC) stripes, or if 
they are centered on rows of oxygen atoms bridging the two neighboring 
metal sites, the so-called bond-centered (BC) stripes, and even 
calculations performed on larger clusters did not yield a definite answer
\cite{Rac05}. Therefore, the purpose of this paper is to study the
stability of both structures based on the mean-field 
method applied either to the extended single-band Hubbard model or the 
single-band Peierls-Hubbard Hamiltonian which includes the so-called 
static phonons \cite{Zaa96}. For stable phases we 
determine charge and magnetization densities, double occupancy, 
kinetic and magnetic energies, and investigate the role of a finite 
electron-lattice coupling.

\section{\label{sec:2} Experimental signatures of stripes }

Experimentally, stripe phases are most clearly detected in insulating 
compounds with a static stripe order, but there is growing evidence 
of fluctuating stripe correlations in metallic and superconducting 
materials. The most direct evidence for stripe phases in doped 
antiferromagnets has come from neutron scattering studies in which charge 
and spin modulations are identified by the appearance of some IC Bragg 
peaks, in addition to those which correspond to the crystal structure. 
However, sometimes sufficiently large crystals are not available for such 
experiments, and one has to resort to other methods capable of probing 
local order. These methods include nuclear magnetic resonance (NMR), 
nuclear quadruple resonance (NQR), muon spin rotation ($\mu$SR),
scanning tunneling microscopy (STM),
and transmission electron microscopy (TEM).
Furthermore, angle-resolved photoemission spectroscopy (ARPES), 
angle-integrated photoemission spectroscopy (AIPES), 
as well as x-ray photoemission (XPS) and 
ultraviolet photoemission (UPS) spectroscopies all provide 
essential information about conspicuous changes in the 
electronic structure when stripe structure sets in. 
Finally, a distinct imprint of the 1D spin-charge modulation 
on transport properties should be detectable as the in-plane anisotropy 
of the resistivity and the Hall coefficient $R_H$.

The abundance of the current evidence on various types of stripe order 
as well as the recent ARPES results on the spectral weight of the cuprate 
superconductors is contained in the review articles by Kivelson 
{\it et al.} \cite{Kiv03}, and by Damascelli {\it et al.} \cite{Dam03}.
Historically, the first compelling evidence for both magnetic and charge 
order in the cuprates was accomplished in a neodymium codoped 
compound La$_{2-x-y}$Nd$_{y}$Sr$_x$CuO$_4$ (Nd-LSCO).    
For $y=0.4$ and $x=0.12$, Tranquada {\it et~al.} \cite{Tra95Nd,Tra96Nd} 
found that the magnetic scattering is not characterized by the 
two-dimensional (2D) AF wave vector ($1/2,1/2$), but by IC peaks 
at the wave vectors $(1/2\pm\epsilon,1/2)$ with $\epsilon = 0.118$. 
Moreover, inspired by the pioneering works demonstrating that the 
staggered magnetization undergoes a phase shift of $\pi$ at the charge 
DWs \cite{Zaa89,Poi89,Sch89,Kat90,Inu91}, the authors found 
additional charge order peaks $(\pm 2\epsilon,0)$, 
precisely at the expected position $2\epsilon=0.236$. Interestingly, 
this doping corresponds to a local minimum in the doping dependence of 
the superconducting temperature $T_c$ in Nd-LSCO \cite{Ich00}, suggesting 
that the static stripes are responsible for this anomalous depression of 
superconductivity. However, it may well be that the apparent correlation 
is entirely accidental and therefore the role of stripes in 
superconductivity remains an open question \cite{Car02}.  

Unfortunately, in early studies Tranquada {\it et~al.} \cite{Tra97Nd} 
detected only magnetic IC peaks at higher doping levels $x=0.15$ and 
$x=0.2$. Nevertheless, systematic NQR studies of Nd-LSCO revealed the 
presence of robust charge stripe order throughout the entire 
superconducting regime of doping $0.07\le x\le 0.25$ \cite{Sin99}. 
Also in a more recent study, both charge and spin superlattice peaks 
at $x=0.15$ were found recently in the neutron diffraction experiments 
by Wakimoto {\it et~al.} \cite{Wak03}.     

In fact, the reason why static stripes could be detected in this compound 
is a structural transition from the low temperature orthorhombic (LTO) to 
the low temperature tetragonal (LTT) phase, induced by the substitution 
for La ions by isovalent Nd ions. This, in turn, provides a pinning 
potential for dynamic stripes and stabilizes the charge order.     
Evidence of a similar pinning potential has also been found both in the 
$\mu$SR and NQR studies of La$_{2-x-y}$Eu$_{y}$Sr$_x$CuO$_4$ (Eu-LSCO)  
with $y\simeq 0.2$ \cite{Kla00,Tei00}. 
Moreover, the connection between the LTT phase and the 
appearance of charge and spin stripe order has been clearly demonstrated 
both in the neutron scattering and $x$-ray diffraction studies on 
La$_{2-x-y}$Ba$_{y}$Sr$_x$CuO$_4$ (Ba-LSCO) with $y=1/8$ \cite{Fuj02Ba,Kim04}.
Finally, static IC charge $(2\pm 2\epsilon,0)$ and magnetic 
$(1/2\pm\epsilon,1/2)$ peaks have been detected within the LTT phase 
of La$_{2-x}$Ba$_{x}$CuO$_4$ (LBCO) with $x=1/8$ \cite{Fuj04}.
The position of the peaks and the established incommensurability 
$\epsilon = 0.118$ are exactly the same as those obtained by Tranquada 
{\it et~al.} \cite{Tra96Nd} for Nd-LSCO. Notably, the peaks that correspond
to charge order appear always at somewhat higher temperature than the 
magnetic ones, indicating that the stripe order is driven by the charge 
instability.

Let us now discuss the experimental evidence of slowly fluctuating stripes 
in La$_{2-x}$Sr$_{x}$CuO$_4$. The main difference between the Ba and Sr 
codoped system is the fact that the latter undergoes a structural phase 
transition from the high-temperature tetragonal (HTT) phase to the LTO 
phase. As a consequence, in the superconducting regime $x\ge0.06$, 
the LSCO system exhibits purely \emph{dynamic} magnetic correlations 
which give rise to IC peaks at the wave vector $(1/2\pm\epsilon,1/2)$ 
specified in tetragonal lattice units $2\pi/a_{tetra}$. In their seminal 
inelastic neutron scattering studies, Yamada {\it et~al.} \cite{Yam98} 
established a remarkably simple relation $\epsilon\simeq x$ for 
$0.06\le x\le 0.12$, followed by a lock-in effect at $\epsilon\simeq 1/8$ 
for larger $x$.

\begin{figure}[t!]
\begin{center}
\includegraphics[width=0.47\textwidth ]{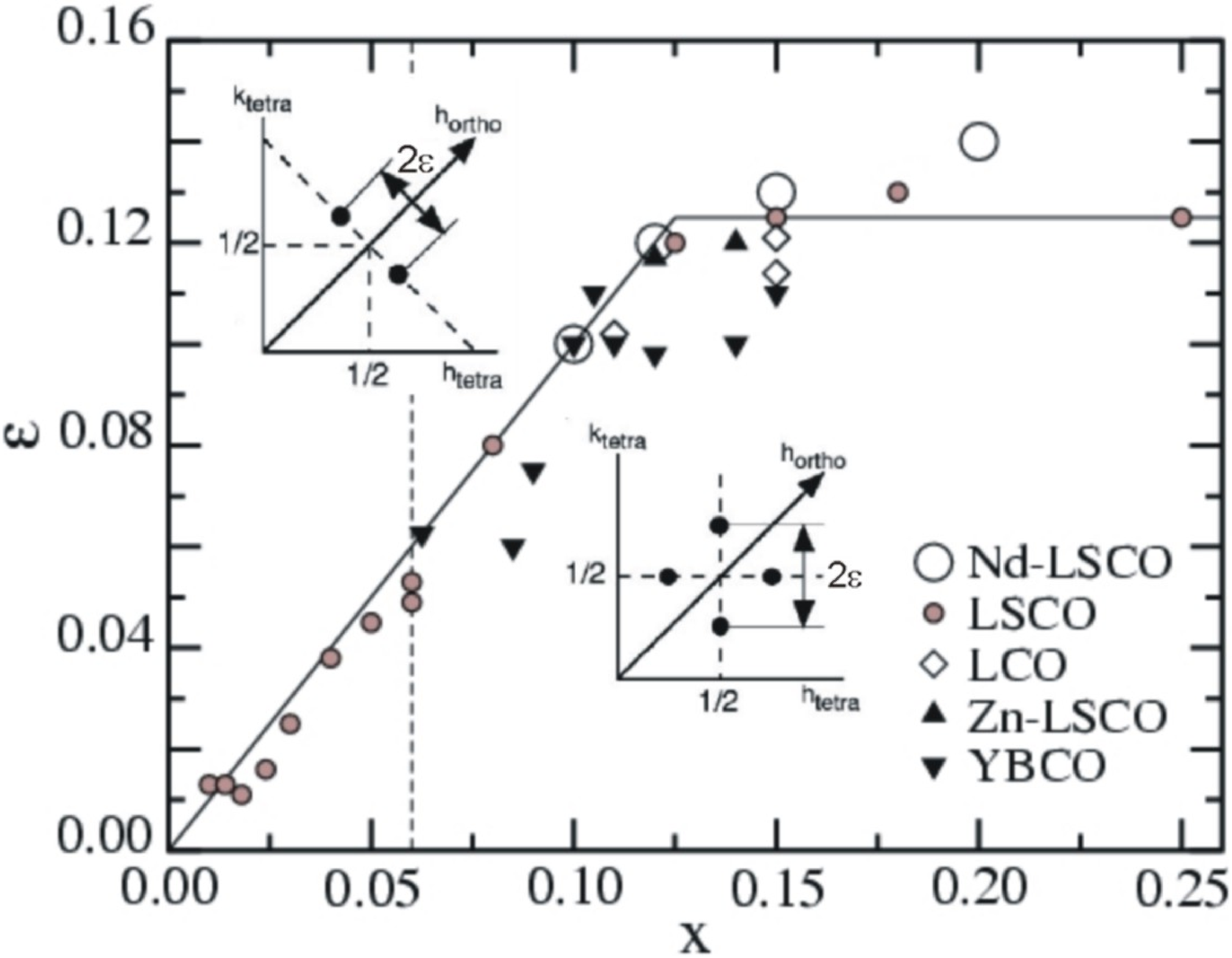}
\end{center}
\caption
{Summary of experimental data illustrating the doping dependence of 
incommensurability $\epsilon$ in the cuprates. 
Results have been obtained by different groups: 
Nd-LSCO (Refs. \cite{Tra95Nd,Tra96Nd,Tra97Nd,Ich00});
LSCO (Refs.~\cite{Yam98,Wak99,Wak00,Fuj02,Mat00,Mat00a,Mat02}); 
LCO (Ref.~\cite{Lee99}); 
Zn-LSCO (Refs.~\cite{Hir98,Kim99}); 
YBCO (Refs.~\cite{Dai01,Moo02}). In 
LSCO, $\epsilon$ has been defined as a distance from the IC peak 
position to the AF wave vector ($1/2,1/2$) either in the orthorhombic 
($x<0.06$) or tetragonal ($x>0.06$) notation (see insets), whereas at 
$x=0.06$, both definitions are used due to the coexistence of diagonal 
and parallel to the Cu-O bonds spin modulations. 
}
\label{fig:epsilon}
\end{figure}

In contrast, in the insulating spin-glass regime of LSCO $x\le 0.06$, 
quasielastic neutron scattering experiments with the main weight at zero 
frequency demonstrate that IC magnetic peaks are located at the wave 
vectors $(1/2\pm\epsilon/\sqrt{2},1/2\pm\epsilon/\sqrt{2})$ 
\cite{Wak99,Wak00,Fuj02}. This phenomenon has often been interpreted 
as the existence of static diagonal stripes, even though no signatures of 
a charge modulation were observed. Another possible explanation is the 
formation of a short ranged spiral order as its chirality also breaks 
the translational symmetry of the square lattice by a clockwise or 
anticlockwise twist \cite{Has04}. Remarkably, even though the spin 
modulation changes from a diagonal to vertical/horizontal one, i.e., 
along Cu-O bonds, at $x$ around 0.06, $\epsilon$ follows the doping $x$ 
reasonably well over the entire range $0.03\le x\le 0.12$, as shown in 
Fig.~\ref{fig:epsilon}. In fact, just for $x=0.06$, both diagonal 
($\epsilon=0.053$) and vertical/horizontal ($\epsilon=0.049$) IC spin 
modulations have been found to coexist \cite{Fuj02}. In a stripe model 
this corresponds to a constant density of 0.5 (0.7) holes per Cu atom in 
the DWs in the vertical/horizontal (diagonal) stripe phases, respectively, 
because of the difference in Cu spacings in the two geometries, i.e., 
$a_{ortho}=\sqrt{2}a_{tetra}$. In contrast, in the narrow region 
$0.02\le x\le 0.024$, IC magnetic peaks are located at the wave vector 
$(1/2\pm\epsilon/2,1/2\pm\epsilon/2)$ with $\epsilon\simeq x$ 
corresponding to a constant charge of one hole/Cu ion along a diagonal DW 
\cite{Mat00,Mat00a,Mat02}. However, below $x=0.02$, this does not hold 
anymore and the incommensurability gets locked with the value 
$\epsilon\simeq0.014$. 

Unfortunately, any concomitant charge ordering has not yet been detected 
in LSCO. Nevertheless, by comparing the data based on the wipeout effect 
of $^{63}$Cu NQR charge order parameter in LSCO with the ones obtained 
from charge stripe compounds as (Nd,Eu,Ba)-LSCO, Hunt {\it et~al.} 
\cite{Hun99} concluded that a similar stripe instability exists in LSCO 
over the whole underdoped superconducting region $1/16\le x\le 1/8$.  
It is also worth mentioning that a very compiling evidence for its 
existence has been established in the measurements of the in-plane 
resistivity and the dynamical infrared conductivity anisotropy 
\cite{And02,Dum03}.   

Experimental detection of IC magnetic peaks in the LTO phase of LSCO 
suggests that the LTT structure is not essential for the appearance of 
stripes. This conjecture has been confirmed in experiments on the 
oxygen doped La$_{2}$CuO$_{4+\delta}$ (LCO) with the orthorhombic 
crystal structure \cite{Lee99}. 
It is also supported by the evidence for \emph{static} 
IC magnetic peaks in another orthorhombic compound  
La$_{2-x}$Sr$_{x}$Cu$_{1-y}$Zn$_y$O$_4$ (Zn-LSCO) 
with $y$ up to $0.03$, even though attempts to observe the charge order 
peaks were unsuccessful \cite{Hir98,Kim99}.
In fact, Zn substitution pins the stripe fluctuations similarly to the 
rare-earth elements. However, in contrast to the latter, it does not 
induce a structural transition to the LTT phase, but provides randomly 
distributed pinning centers that promote meandering of stripes and 
correspondingly broadens IC peaks. 

An important question is whether charge stripes appear solely in 
monolayered lanthanum compounds or if they are a generic feature of all 
the cuprates. The latter conjecture seems to be supported by inelastic 
neutron scattering experiments on bilayered 
YBa$_{2}$Cu$_{3}$O$_{6+\delta}$ (YBCO) compounds that have identified the 
presence of IC spin fluctuations throughout its entire superconducting 
regime \cite{Dai01}. In fact, as the doped charge is nontrivially 
distributed between the CuO$_2$ planes and CuO chains, it is very 
difficult to determine the precise doping level $x$ in the CuO$_2$ sheet 
of YBCO. Nevertheless, systematic studies by Dai {\it et al.} \cite{Dai01} 
have shown that the incommensurability in YBCO increases initially with 
doping but it saturates faster than in LSCO, i.e., already at 
$x\simeq 0.1$ with the value $\epsilon\simeq 0.1$. Unfortunately, there 
is no any compelling explanation that would account for such a different 
behavior of $\epsilon$ in both systems. 
Eventually, charge order peaks have been observed in YBCO$_{6.35}$  
but in spite of several attempts, no static charge order could be 
detected in YBCO$_{6.5}$ and YBCO$_{6.6}$ so far \cite{Moo02}.

Furthermore, although some neutron scattering experiments have been 
performed on Bi$_2$Sr$_2$CaCu$_2$O$_{8+\delta}$ (BSCCO) sample, the 
sample has only produced weak evidence of the IC structure \cite{Moo97}. 
In contrast, Fourier transform of the recent STM data has revealed some 
IC peaks corresponding to a four-period modulation of the local density 
of states along the Cu-O bond direction, which may imply the existence of 
stripes \cite{How03}. Nevertheless, definite answer pertinent to the 
appearance of stripes in all the cuprates remains still unsettled and 
further experiments are required to reach an unambiguous conclusion, even 
though the summary of the experimental data illustrating the doping 
dependence of the incommensurability $\epsilon$ in cuprates, depicted in 
Fig.~\ref{fig:epsilon}, includes an array of compounds.

Tendency towards phase separation is also a starting point to understand 
the doping evolution of the electronic structure in LSCO and Nd-LSCO. 
For example, ARPES spectra measured at the $X=(\pi,0)$ point in LSCO 
show that even though the data are solely characterized by a single high 
binding energy feature in the insulating regime, upon increasing 
doping one observes a systematic transfer of spectral weight from the 
high- to the low binding energy part \cite{Ino00}. Consequently, a 
well-defined quasiparticle (QP) peak develops near the optimal doping. 
In contrast, the intensity near the $S=(\pi/2,\pi/2)$ point remains 
suppressed for the entire underdoped regime so that a QP peak is 
observed only for $x\ge 0.15$.  

Another peculiar feature of the ARPES band dispersion is extensively 
discussed in the literature saddle point at the $X$ point, the so-called 
flat band \cite{Ino02}. As hole doping increases, the flat band moves 
monotonically upwards and crosses the Fermi level $E_F$ at $x\simeq 0.2$. 
This is reflected in the enhancement of the DOS at the chemical potential 
$N(\mu)$ observed by AIPES \cite{Ino98}. 

The experimental distribution of the photoemission spectral weight near 
the $X$ and $S$ points in doped LSCO has been nicely reproduced using the 
DMFT approach for vertical SC stripes obtained within the Hubbard model 
\cite{Fle00}. As a consequence of the stripe order, the obtained spectra 
along the $\Gamma-X-M$ path were not equivalent to those along the 
$\Gamma-Y-M$ one, with $\Gamma=(0,0)$ and $Y=(0,\pi)$. Moreover, as in 
the experiment, the spectral weight along the $\Gamma-X$ direction was 
suppressed close to the $\Gamma$ point and simultaneously enhanced at the 
$X$ point. Furthermore, in the framework of stripes, the flat QP band 
near the $X$ point with a large intensity at the maximum below the 
chemical potential $\mu$ follows from a superposition of the 
dispersionless 1D metallic band along the $x$ direction, formed by holes 
propagating along the vertical domain walls, and an insulating band that 
stems from the AF domains. In contrast, an AF band at the $Y$ point is 
characterized by a high binding energy well below $\mu$ and consequently 
the spectral weight at $\omega=\mu$ almost vanishes. Finally, a distinct 
gap for charge excitations should open at $\mu$ near the $S$ point. This 
gap follows indeed from the stripe structure --- while the system may be 
metallic along the stripes, i.e., in the antinodal directions $\Gamma-X$ 
or $\Gamma-Y$, the low-energy excitations should be noticeably 
suppressed along the nodal direction $\Gamma-S$ crossing all the stripes. 
This conjecture is also supported either by the ED studies \cite{Toh99} 
or by the analytical approach based on variational trial wave function 
within the string picture \cite{Wro00}, both applied to the 
$t$-$t'$-$t''$-$J$ model, or by the CPT for the $t$-$J$ model \cite{Zac00}.

In fact, the low-energy spectral weight of Nd-LSCO at $x=0.12$, a model 
compound for which the evidence of spin and charge stripe order is 
the strongest, is also mostly concentrated in flat regions along the 
$\Gamma-X$ and $\Gamma-Y$ directions, while there is only little spectral 
weight along the $\Gamma-S$ direction \cite{Zho99}. On the other hand, 
ARPES spectra of both LSCO and Nd-LSCO at $x=0.15$ have revealed not only 
the presence of flat bands around the $X$ and $Y$ points, but also the 
existence of appreciable spectral weight at $E_F$ in the nodal region 
\cite{Zho01}. While the observation of flat segments might be directly 
ascribed to 1D domain walls \cite{Sal96}, detection of nodal spectral 
weight poses a formidable task to develop a theory that would describe 
the electronic structure resembling the FS of a fully 2D system because, 
as it was already stressed out, the nodal spectral weight is expected to 
be suppressed in a static SC stripe picture 
\cite{Fle00,Toh99,Wro00,Zac00}. Indeed, the experimentally established 
FS looks rather like the one arising from disorder or from dynamically 
fluctuating stripes \cite{Sal96}.   
 
Alternatively, guided by the CPT results showing that while the SC 
stripes yield little spectral weight near the nodal region, the BC ones 
reproduce quite well the nodal segments \cite{Zac00}, Zhou {\it et~al.} 
\cite{Zho01} have conjectured that the experimental FS may result from 
the coexistence of the SC and BC stripes. Within this framework, upon 
increasing doping the BC stripes are formed at the expense of the SC 
ones. This scenario is particularly interesting because it has been 
shown that the BC stripe, in contrast to its SC counterpart, enhances 
superconducting pairing correlations \cite{Eme97}. 
The relevance of a bond order at the doping level $x=0.15$ is supported 
by recent studies of the ARPES spectra in a system with the BC stripes 
\cite{Wro05}. These studies have yielded pronounced spectral weight both 
in the nodal and antinodal directions, reproducing quite well the 
experimental results in Nd-LSCO and LSCO \cite{Zho01}. Furthermore, the 
stripe scenario would also explain the origin of the already discussed 
two components seen in the ARPES spectra at the $X$ point near $x=0.05$ 
\cite{Ino00}. Indeed, the response from the AF insulating regions would 
be pushed to the high binding energies due to the Mott gap, whereas the 
charge stripes would be responsible for the other component near $E_F$. 

Existence of DWs should also give rise to the appearance of new states 
inside the charge-transfer gap that would suppress the shift of the 
chemical potential $\mu$ in the underdoped regime $x<1/8$ where 
$\epsilon$ increases linearly. Such pinning of $\mu$ in LSCO was indeed 
deduced from XPS experiments \cite{Ino97}. In contrast, in the overdoped 
region with a lock-in effect of $\epsilon$, the number of stripes per
unit cell saturates, doped holes penetrate into the AF domains, and 
consequently $\mu$ would move fast with doping in agreement with the 
experimental data. The picture of broadened stripes and holes spreading 
out all over the AF domains above $x=1/8$ is also indicated by the doping 
dependence of the resistivity and the Hall coefficient $R_H$ in Nd-LSCO. 
Namely, a rapid decrease in the magnitude of $R_H$ for doping level 
$x\le 1/8$ at low temperature provides evidence for the 1D charge 
transport, whereas for $x>1/8$, relatively large $R_H$ suggests a 
crossover from the 1D to 2D charge transport \cite{Nod99}. Altogether, 
it appears that the metallic stripe picture does capture the essence of 
the low-lying physics for Nd-LSCO and LSCO systems.    

Conversely, it is important to note that so far no evidence of IC peaks 
has been detected in any electron-doped cuprates superconductors. 
Instead, the neutron scattering experiments have established only 
\emph{commensurate} spin fluctuations as in Nd$_{2-x}$Ce$_x$CuO$_{4}$ 
(NCCO), both in the superconducting and in normal state \cite{Yam03}. 
Moreover, observation of such peaks is consistent with the XPS 
measurements in NCCO showing that the chemical potential increases
monotonously with electron doping \cite{Har01}.

\section{\label{sec:3} Numerical results}     

In this Section we attempt a systematic investigation of the properties 
and relative stability of filled vertical and diagonal stripes. We shall 
see that in spite of the difficulty to stabilize the ground state with 
half-filled stripes (one hole per every two atoms in a DW), the 
mean-field framework is useful as providing a generic microscopic 
description of filled inhomogeneous reference structures with the 
filling of one doped hole per stripe unit cell. Their special stability 
rests on a gap that opens in the symmetry broken state between the 
highest occupied state of the lower Hubbard band and the bottom of the 
so-called mid-gap bands, i.e., some additional unoccupied bands lying 
within the Mott-Hubbard gap that are formed due to holes propagating 
along DWs \cite{Zaa96}.

Here, we extend early HF studies of the filled DWs 
\cite{Poi89,Sch89,Kat90,Inu91} and determine a phase diagram of the 
Hubbard model with an anisotropic nearest-neighbor hopping $t$ by 
varying the on-site Coulomb repulsion $U$ and investigating locally 
stable structures for representative hole doping levels $x=1/8$ and 
$x=1/6$. We also report the changes in stability of the stripe structures 
in the extended Hubbard model due to the next-neighbor hopping $t'$ and 
to the nearest-neighbor Coulomb interaction $V$. Finally, in order to 
gain a comprehensive understanding of the competition between different 
types of stripes in a realistic model, we include lattice degrees of 
freedom induced by a static Peierls electron-lattice coupling. 

\subsection{\label{sec:3a} Extended single-band Hubbard model}

The starting point for the analysis of stripe structures is the extended 
single-band Hubbard model, which is widely accepted as the generic model 
for a microscopic description of the cuprate superconductors \cite{Fei96},
\begin{equation}
H=-\sum_{ij\sigma}t^{}_{ij}c^{\dag}_{i\sigma}c^{}_{j\sigma} +
  U\sum_{i}n^{}_{i\uparrow}n^{}_{i\downarrow} +
  V\sum_{\langle ij\rangle}n^{}_in^{}_j,
\label{eq:Hubb}
\end{equation}
where the operator $c^{\dag}_{i\sigma}$ $(c^{}_{j\sigma})$ creates
(annihilates) an electron with spin $\sigma$ on lattice site $i$ ($j$),
and $n_i=c^{\dag}_{i  \uparrow}c^{}_{i  \uparrow}
       + c^{\dag}_{i\downarrow}c^{}_{i\downarrow}$
gives the electron density. The hopping $t_{ij}$ is $t$ on the bonds
connecting nearest neighbors sites $\langle i,j\rangle$ and $t'$ for 
second-neighbor sites, while the on-site and nearest-neighbor Coulomb 
interactions are, respectively, $U$ and $V$. 

The model can be solved self-consistently in real space within the HF, 
where the interactions are decoupled into products of
one-particle terms becoming effective mean fields that act on each 
electron with the same strength. This approximation basically involves 
solving an eigenvalue problem. The obtained wavefunctions form a new 
potential and hence the Hamiltonian for a new eigenvalue problem. Typically, 
the new potential is chosen as some linear combination of the current 
and preceding potential. The iterations are continued until the input and 
output charge density and energy do not change within some prescribed 
accuracy. The most significant drawback of this method is that it neglects 
correlations. Electron correlation changes the system properties and 
manifests itself in the decrease of the ground state energy. The difference 
between the energy of the exact ground state and the energy obtained within 
the HF is thus called the correlation energy. It arises from the fact that 
an electron's movement is correlated with the electrons around it, 
and accounting for this effect lowers further the energy, beyond the 
independent electron approximation. 

\begin{figure}[!t]
\begin{center} 
\unitlength=0.01\textwidth
\begin{picture}(100,46)
\put(3,46){\includegraphics[scale=0.25,angle=270]{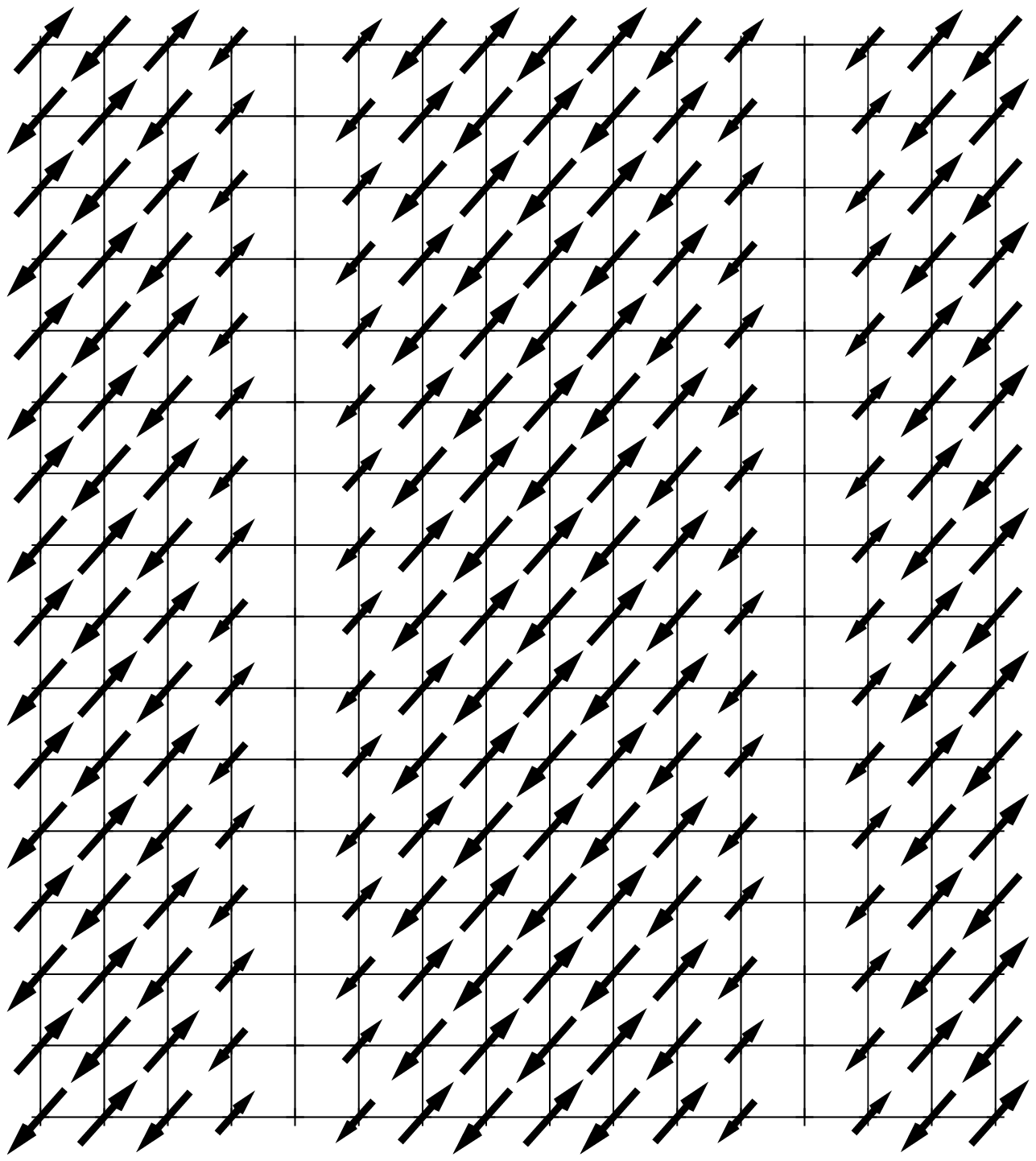}}
\put(25,46){\includegraphics[scale=0.25,angle=270]{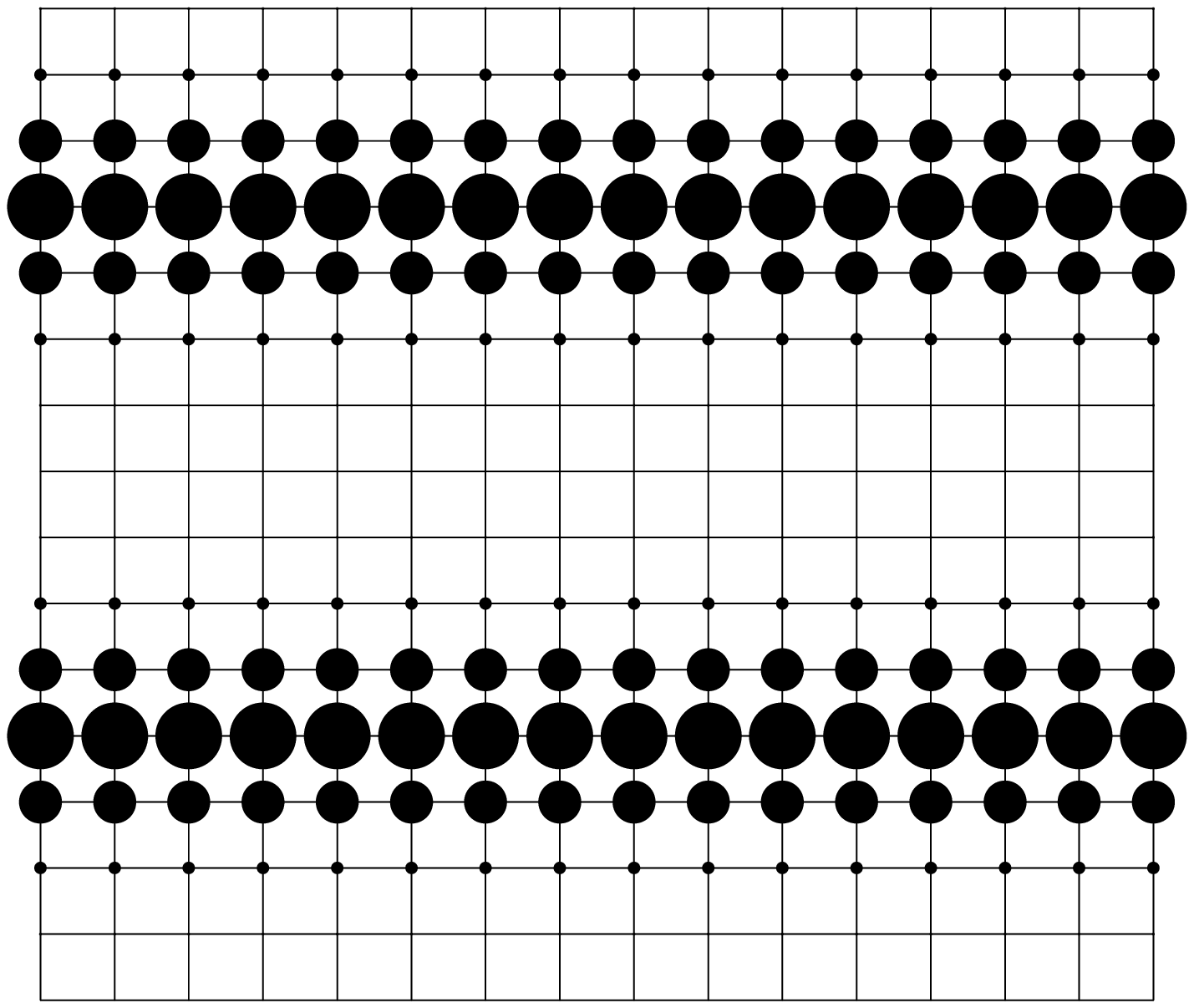}}
\put(3,24){\includegraphics[scale=0.25,angle=270]{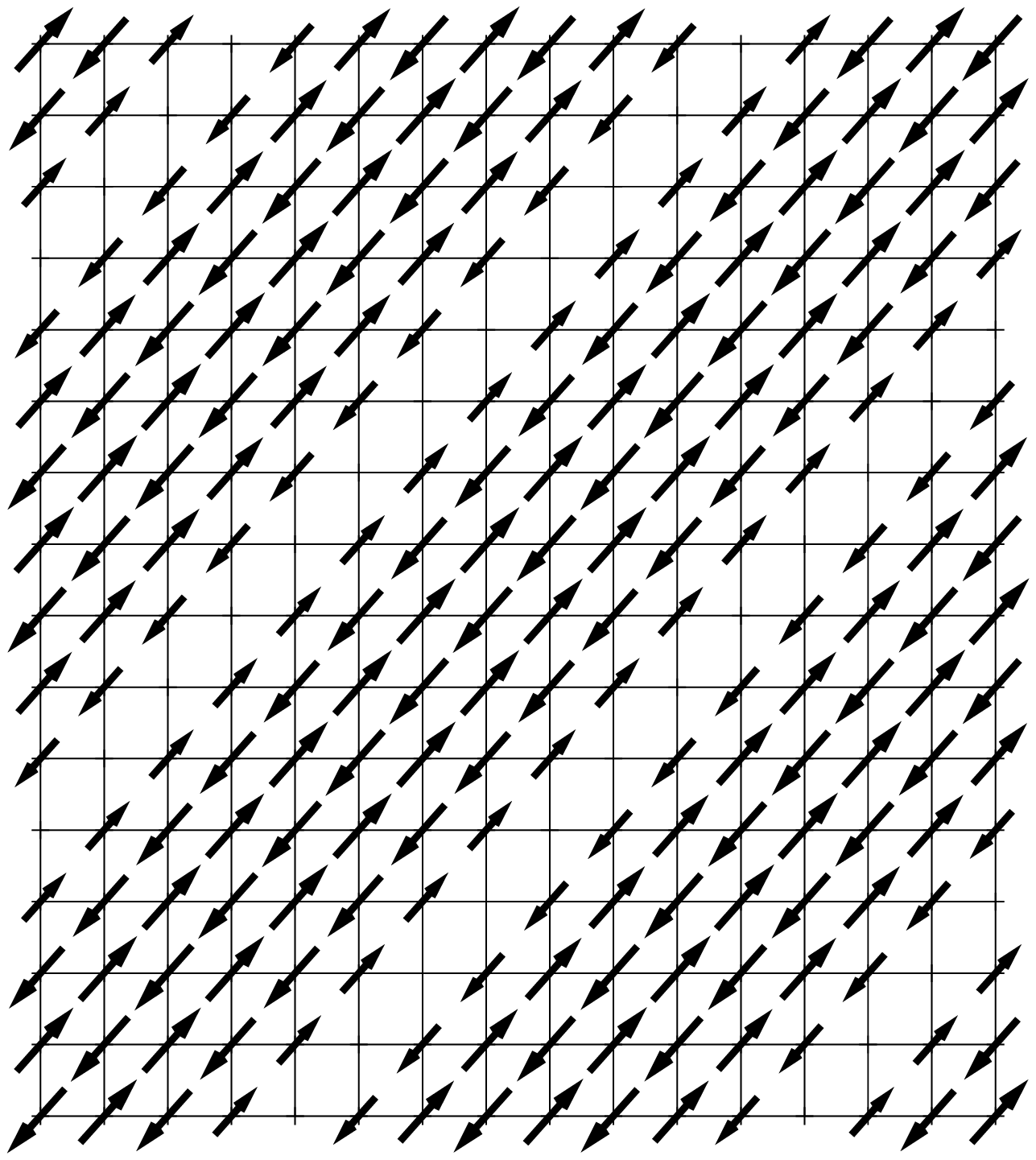}}
\put(25,24){\includegraphics[scale=0.25,angle=270]{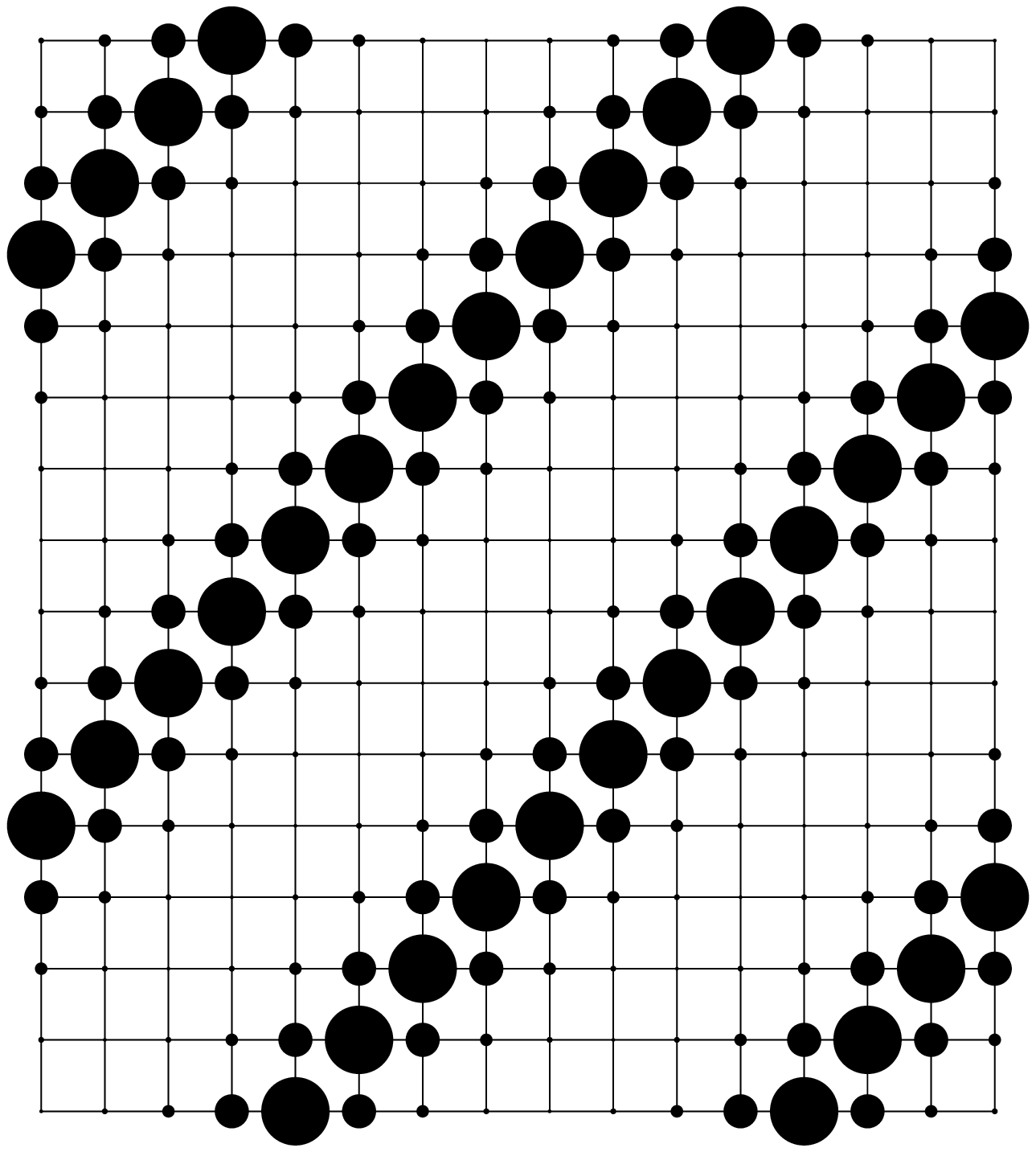}}
\put(2.5,0){\vector(1,0){10}}
\put(2.5,0){\vector(0,1){10}}
\put(5,1){$l_x$}
\put(0.5,4){$l_y$}
\end{picture}
\end{center}
\caption{ Vertical site-centered (VSC) and 
diagonal site-centered (DSC) 
stripe phases as found for $U/t=5$ at hole doping $x=1/8$. 
The length of arrows is proportional to the magnetization 
$\langle S_i^z\rangle$ and the hole density $\langle n_{{\rm h}i}\rangle$ 
is scaled by the diameter of black circles.}
\label{fig:16SC}
\end{figure}

We do not consider noncollinear spin configurations, and use the most 
straightforward version of the HF with a product of two 
separate Slater determinants for up and down spins, whence,
\begin{equation}
 n_{i\uparrow}n_{i\downarrow}\simeq
n_{i\uparrow}\langle n_{i\downarrow}\rangle
 +\langle n_{i\uparrow}\rangle n_{i\downarrow}
 -\langle n_{i\uparrow}\rangle\langle n_{i\downarrow}\rangle.
\label{eq:MF}
\end{equation}
A similar decoupling is performed for the nearest-neighbor Coulomb
interaction. Calculations were performed on $12\times 12$ ($16\times 16$)
clusters for $x=1/6$ ($x=1/8$) with periodic boundary conditions, and we
obtain stable stripe structures with AF domains of width five atoms for
$x=1/6$ and seven atoms for $x=1/8$. Typical solutions at $x=1/8$ are
shown in Fig.~\ref{fig:16SC} with the local hole density,
\begin{equation}
\langle n_{{\rm h}i}\rangle = 
1-\langle n_{i\uparrow} + n_{i\downarrow}\rangle,
\label{eq:nhi}
\end{equation}
scaled by the diameter of the black circles and the length of the 
arrows being proportional to the amplitude of local magnetization density,
\begin{equation}
\langle S_i^z\rangle = 
\tfrac{1}{2}|\langle n_{i\uparrow} - n_{i\downarrow}\rangle|.
\label{eq:Szi}
\end{equation}

These structures possess nonmagnetic DWs with enhanced hole 
density which separate AF domains having hole density almost unchanged 
with respect to the undoped case. Note that the AF sites on each side of 
the DWs have a phase shift of $\pi$. 

In order to appreciate better the microscopic reasons 
of such arrangement let us consider a small cluster consisting of
three atoms filled by two electrons and one hole (with respect to 
half-filling with the electron density $n=1$ per site). For simplicity 
we assume that the electrons are confined to the considered cluster 
owing to large Coulomb interaction $U\gg t$, and we do not
take into account any interactions with the AF background. There are two
possible candidates for the ground state. The first one corresponds to a 
hole added to three atoms of a single AF domain in which, if we suppose 
that a $\downarrow$-spin electron is replaced by a hole, the two remaining 
$\uparrow$-spin electrons can be found in one of three allowed 
configurations: $\{\uparrow, 0, \uparrow\}$, $\{\uparrow, \uparrow, 0\}$, 
and $\{0,\uparrow,\uparrow\}$ (the other configurations are excluded by 
the Pauli principle). Hence, this polaronic state gives the total energy,
\begin{equation}
E_P=-\sqrt{2}t,
\label{eq:pol}
\end{equation}
and the Coulomb interaction $U$ does not contribute.

\begin{figure}[!t]
\begin{center} 
\unitlength=0.01\textwidth
\begin{picture}(100,46)
\put(3,46) {\includegraphics[scale=0.25,angle=270]{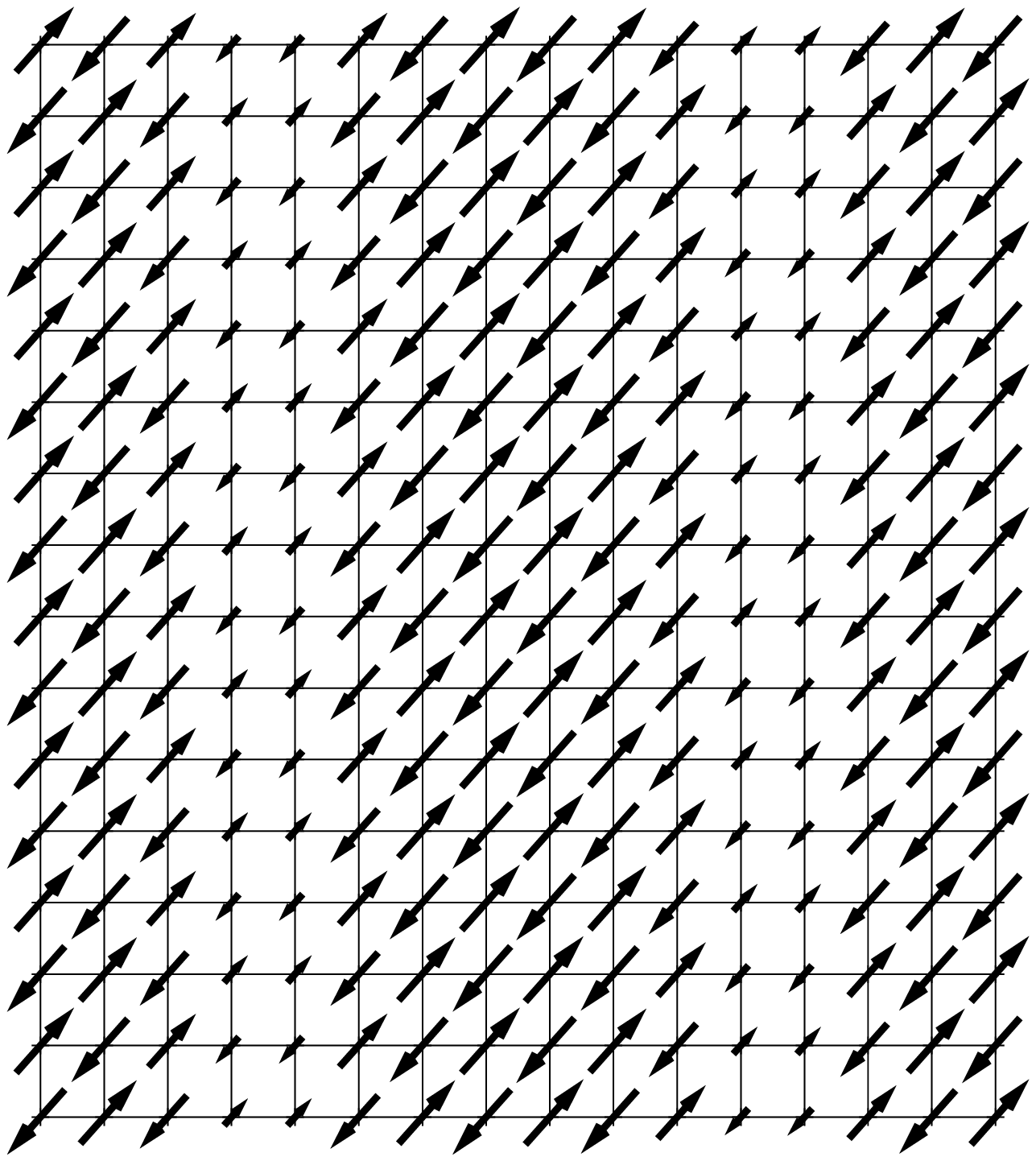}}
\put(25,46) {\includegraphics[scale=0.25,angle=270]{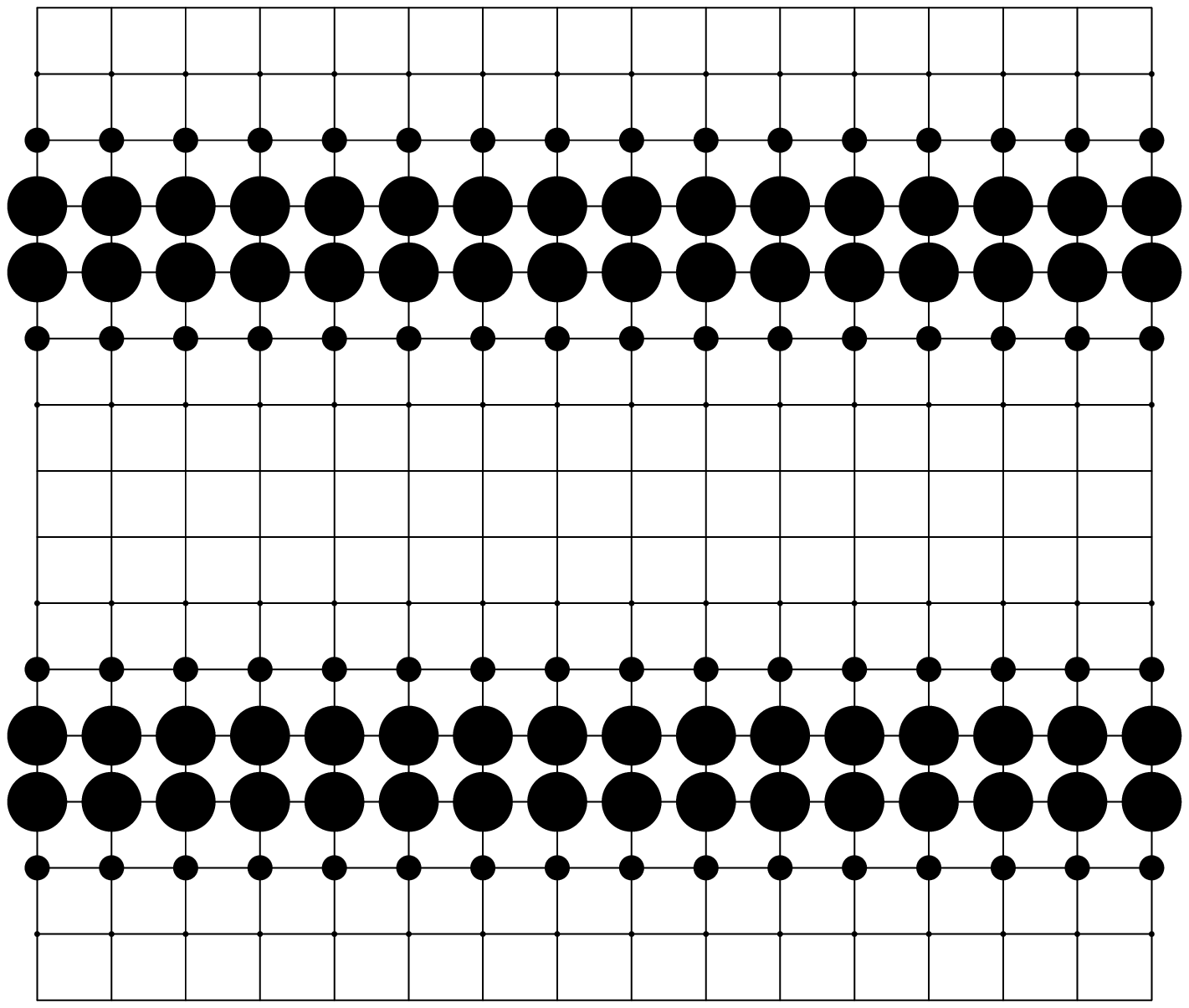}}
\put(3,24){\includegraphics[scale=0.25,angle=270]{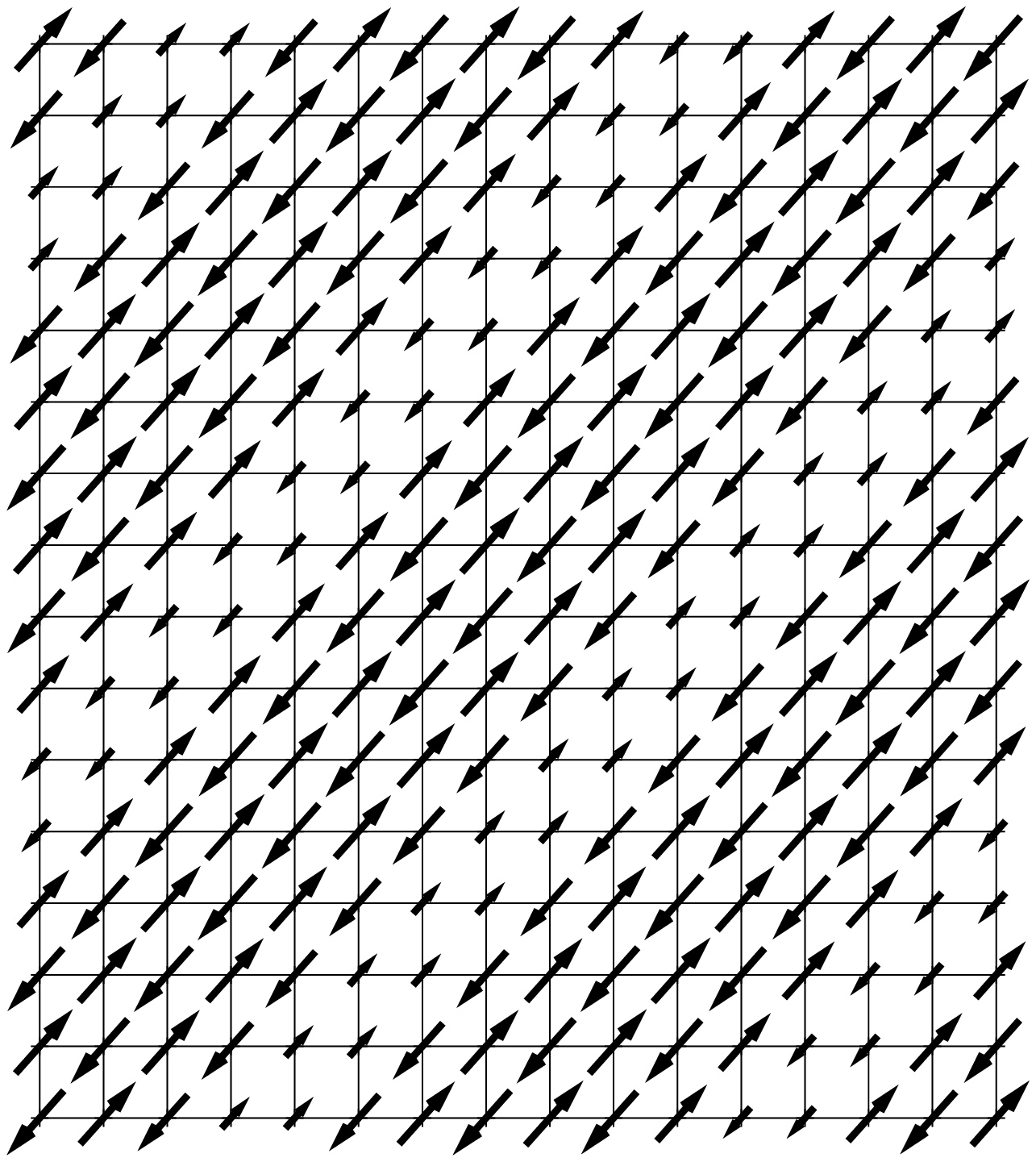}}
\put(25,24){\includegraphics[scale=0.25,angle=270]{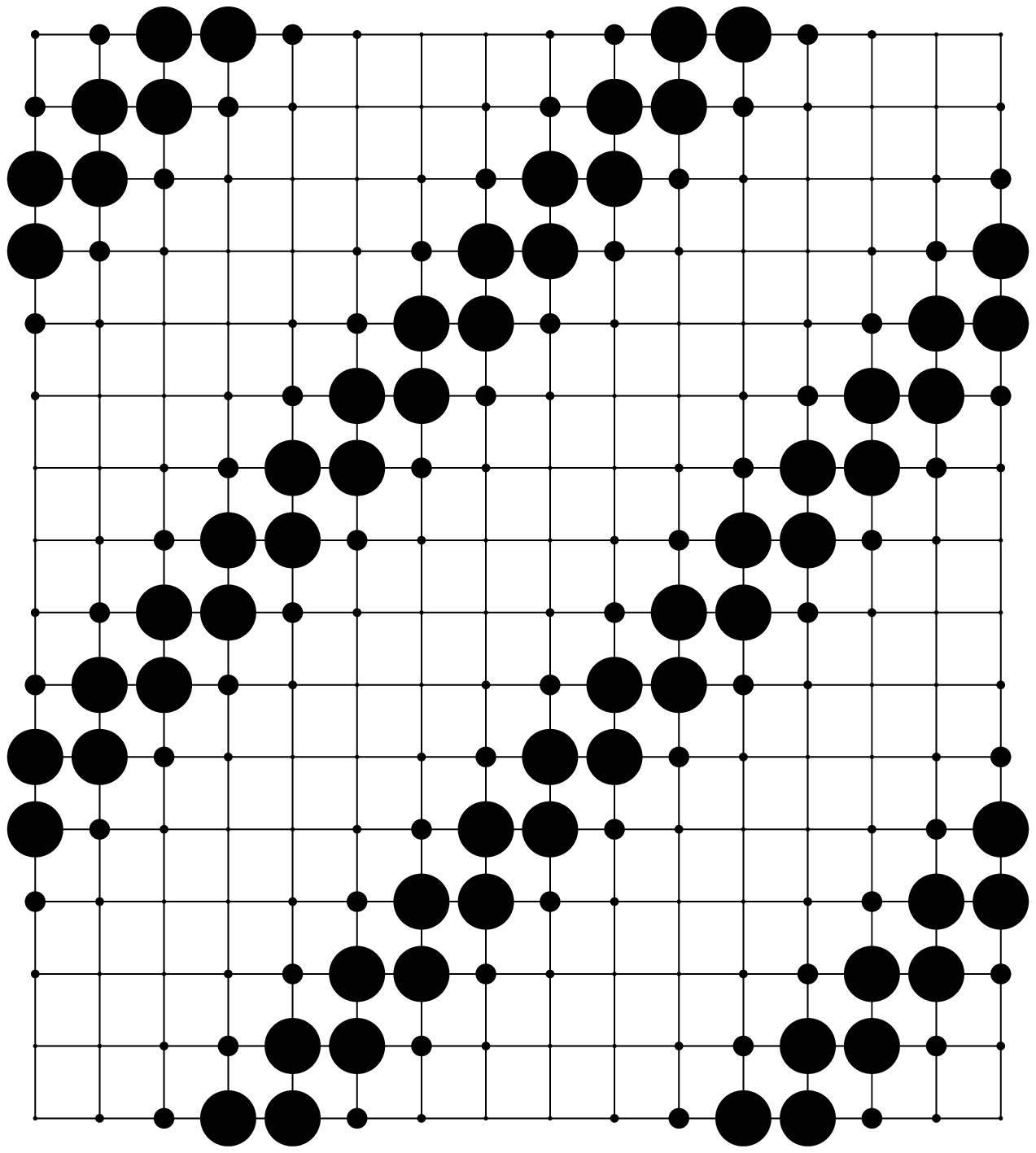}}
\put(2.5,0){\vector(1,0){10}}
\put(2.5,0){\vector(0,1){10}}
\put(5,1){$l_x$}
\put(0.5,4){$l_y$}
\end{picture}
\end{center}
\caption{ Vertical bond-centered (VBC) 
and diagonal bond-centered (DBC) 
stripe phases as found for $U/t=5$ at hole doping $x=1/8$. 
The meaning of the arrows and black circles as in Fig.~\ref{fig:16SC}.}
\label{fig:16BC}
\end{figure}

A different situation is obtained when a hole occupies instead a DW 
separating two AF domains. Delocalization leads then to similar three 
configurations to those obtained above: 
$\{\uparrow, 0 ,\downarrow\}$, $\{\uparrow, \downarrow, 0\}$, and 
$\{0, \uparrow, \downarrow\}$, but in addition, three configurations with 
one doubly occupied site $\{\uparrow\downarrow, 0, 0\}$,
$\{0, \uparrow\downarrow, 0\}$, and $\{0, 0, \uparrow\downarrow\}$, can be 
reached as excited states which cost Coulomb energy $U$. Moreover, three 
other configurations with interchanged $\uparrow$- and $\downarrow$-
spins are then also accessible via the decay of double occupancies: 
$\{\downarrow, 0, \uparrow\}$, $\{\downarrow, \uparrow, 0\}$, and 
$\{0, \downarrow, \uparrow\}$. In the regime of large $U$, the total 
energy in the ground state can be found in a perturbative way, and as 
a result one obtains,
\begin{equation}
E_S=-\sqrt{2}t - \frac{4t^2}{U}.
\label{eq:sol}
\end{equation}
Therefore, the Hilbert space for the latter solitonic solution is larger 
and one finds that this solution is always more stable than the polaronic 
one \cite{Zaa96}. The argument applies also to 2D systems, where the DWs 
are more stable than the lines of polarons in an AF background.

We compare the stability of such nonmagnetic SC domain walls with 
the BC stripe phases in which DWs are formed by pairs of magnetic atoms, 
as obtained by White and Scalapino \cite{Whi98} 
(\textit{cf}. Fig.~\ref{fig:16BC}). In the three-band model, SC (BC) 
stripes correspond to DWs centered at metal (oxygen) sites, 
respectively \cite{Miz97,Yu98,Sad00,Lor02}.     

\subsection{\label{sec:3b} Effect of hopping anisotropy}

We begin by setting $t'=0$ and $V=0$ with the goal of elucidating the 
effects of hopping anisotropy on the stripes. This is motivated by the 
fact that the first detection of static stripes in both charge and spin 
sectors was accomplished in Nd-LSCO \cite{Tra95Nd} indicating that 
rare-earth elements doping is in some way helpful for pinning the stripe 
structure. Indeed, it produces a structural transition in the system
from the LTO to LTT phase \cite{Buc94}. Both phases involve a distortion
of the CuO$_2$ plane by rotation of the CuO$_6$ octahedra. In the 
LTO phase the tilt axis runs diagonally within the copper plane,
such that all the oxygen atoms are displaced out of the plane. Conversely,
in the LTT phase this rotation takes place around an axis oriented 
along the planar Cu$-$O bonds. Therefore, oxygen atoms on the tilt 
axis remain in the plane, while the ones in the perpendicular direction 
are displaced out of the plane. This provides a microscopic origin
for in-plane anisotropies --- the Cu$-$Cu hopping amplitude $t$ depends 
on the Cu$-$O bond and it is isotropic in the LTO phase and anisotropic 
in the LTT one. For a physical tilt angle of order 5$^{\circ}$, the 
relative anisotropy taking $t_y<t_x$, 
\begin{equation}
\epsilon_t = \frac{|t_x-t_y|}{t_y}, 
\label{eq:et}
\end{equation}
is weak and amounts to $\epsilon_t=1.5\%$ \cite{Nor01,Kam01}. The 
direction with a larger hopping amplitude coincides with the direction 
of a stronger superexchange coupling $J$.

\begin{figure}[t!]
\begin{center}
\includegraphics[width=0.47\textwidth ]{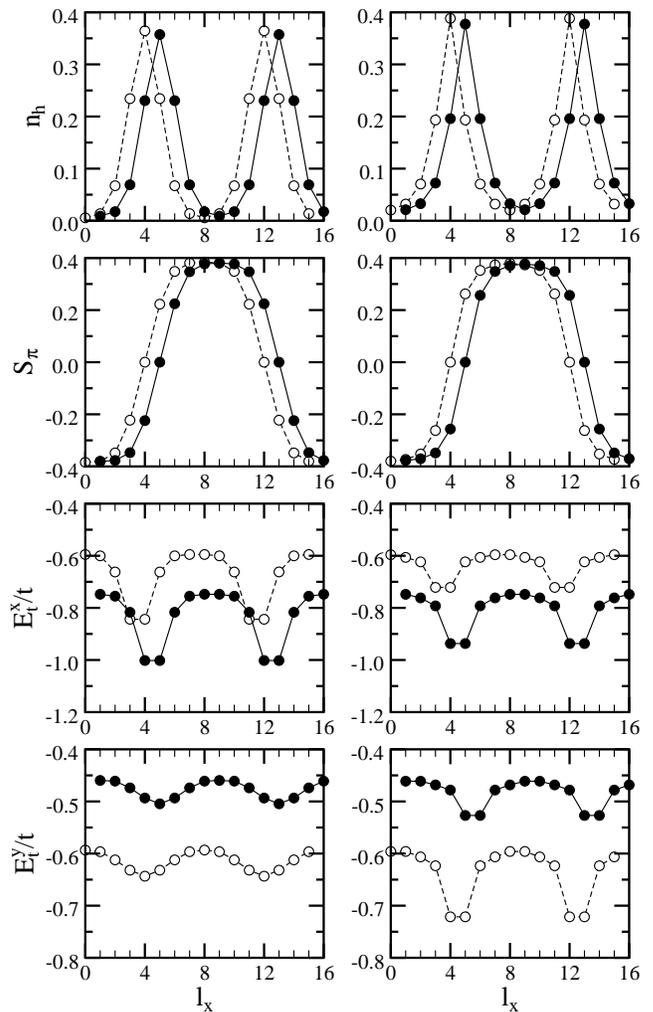}
\end{center}
\caption
{ Local hole $n^{}_{\rm h}(l_x)$ (top) and magnetization $S_{\pi}(l_x)$ 
(second row) density; kinetic energy $E_{t}^{x}(l_x)$ (third row) and 
$E_{t}^{y}(l_x)$ (bottom) projected on the bonds in the $x$-($y$)-directions, 
respectively, of the VSC (left) and DSC (right) stripe phases shown in 
Fig.~\ref{fig:16SC} (open circles) as well as of the ones obtained in the 
anisotropic model with $t_x/t_y=1.22$ (filled circles). For clarity, the 
latter are shifted by one lattice constant from the origin of the coordinate 
system.  
}
\label{fig:SC16}
\end{figure}
The possible relationship between this anisotropy and the onset of stripe 
phases has been intensively studied within anisotropic Hubbard 
($t_x\ne t_y$) or $t$-$J$ ($t_x\ne t_y$, $J_x\ne J_y$) models by means of 
various techniques: unrestricted HF approach \cite{Nor01}, DMRG 
\cite{Kam01}, and QMC method \cite{Bec01}. 
The in-plane anisotropies might also be represented theoretically by 
on-site potentials as in the QMC study by Riera \cite{Rie01}. 
All these studies have shown a pronounced tendency to forming stripe 
phases, which manifests itself by the reduction of their energy 
\cite{Nor01,Kam01}, accompanied by the appearance of IC peaks in the spin 
and charge structure factor \cite{Bec01,Rie01}. It appears that
a finite anisotropy of the next-nearest hopping term $t'$ might play 
a role in stabilizing diagonal incommensurate peaks observed in the 
spinglass phase of LSCO ($0.02\leqslant x\leqslant 0.06$) 
\cite{Wak99,Wak00, Mat00, Mat00a,Fuj02,Mat02}. Indeed, although the LTO 
phase is usually considered as isotropic, which is the case for 
nearest-neighbor hopping and interaction, a different length of the 
orthorhombic axes implies the need for an anisotropic $t'$ parameter. 
Exact diagonalization studies incorporating such anisotropy have shown 
that it strongly strengthens hole correlations along one direction and 
suppresses them along the other, resulting in a 1D pattern of holes 
\cite{Tip03}.

\begin{figure}[t!]
\begin{center}
\includegraphics[width=0.47\textwidth ]{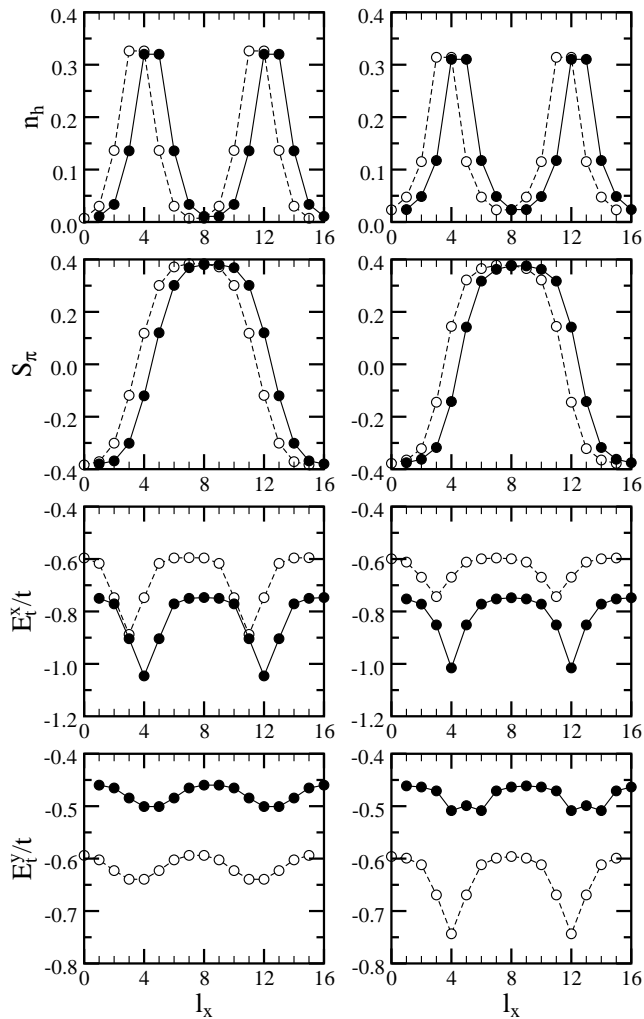}
\end{center}
\caption
{
The same as in Fig.~\ref{fig:SC16} but for the BC stripe phases.
}
\label{fig:BC16}
\end{figure}

It turns out, however, that the variation of the hopping anisotropy 
$\epsilon_t$ (\ref{eq:et}) has only a little visible effect on the local 
hole density, 
\begin{equation}
n_{\rm h}(l_x) = 1-\langle n_{(l_x,0),\uparrow} + 
n_{(l_x,0),\downarrow}\rangle, 
\label{eq:nh}
\end{equation}
shown in Fig.~\ref{fig:SC16} as a function of the $x$-direction 
coordinate $l_x$ for a given $y$-direction coordinate $l_y=0$, even at 
the unrealistically large anisotropy level $\epsilon_t=22\%$, 
corresponding to $t_x/t=1.1$ and $t_y/t=0.9$. Similarly, the anisotropy 
does not modify the modulated magnetization density,
\begin{equation}
S_{\pi}(l_x) = (-1)^{l_x}\tfrac{1}{2}
\langle n_{(l_x,0),\uparrow} - n_{(l_x,0),\downarrow}\rangle,
\label{eq:Spi}
\end{equation}
with a site dependent factor $(-1)^{l_x}$ compensating modulation
of the staggered magnetization density within a single AF domain. 

In contrast, the strong effect of finite anisotropy $\epsilon_t$ 
(\ref{eq:et}) is clearly demonstrated by variation of the expectation 
values of the bond hopping terms along the $x$- and $y$-directions,  
\begin{align}
E_t^x(l_x) &= -t_x\sum_{\sigma}
\bigl\langle c_{(l_x,0),\sigma}^{\dag}c_{(l_x+1,0),\sigma}^{} + 
h.c\bigr\rangle,
\label{eq:Etx}\\
E_t^y(l_x) &= -t_y\sum_{\sigma}
\bigl\langle c_{(l_x,0),\sigma}^{\dag}c_{(l_x,1),\sigma}^{} + 
h.c\bigr\rangle.
\label{eq:Ety}
\end{align}
These features are seen in Fig.~\ref{fig:SC16}. For the VSC stripes one 
finds a large anisotropy in the values of the kinetic energies 
(\ref{eq:Etx}) and (\ref{eq:Ety}), which becomes especially pronounced 
beside the stripes, and is strongly reinforced by the hopping anisotropy. 
Therefore, taking into account that the hopping between two different 
charge densities is favored over motion between equal densities, one 
should expect that transverse charge fluctuations will always tune the 
direction of DWs along the weaker hopping direction in the anisotropic 
model. Analogous conclusion based on Fig.~\ref{fig:BC16} might be drawn 
concerning the orientation of the VBC stripes. 

Regarding diagonal stripes, although a finite anisotropy in hopping is   
also reflected in the kinetic energy anisotropy, a system with either the
DSC or DBC stripe pattern becomes topologically frustrated and 
consequently may gain less kinetic energy compared to a system with 
vertical stripes, taking a full advantage of the hopping anisotropy 
(\textit{cf}. Tables~\ref{tab:16E} and \ref{tab:16Ean}).

\begin{table}[!t]
\begin{center}
\begin{tabular}{ccccccc}
\hline\hline
&            &  $E_t^x/t$  &  $E_t^y/t$  & $E_U/t$ &  $E_{\textrm{tot}}/t$  \\
\hline
&{\bf VB(S)C}& $-$0.6753   & $-$0.6147  &  0.4900  & $-$0.8000   \\   
&{\bf DBC}   & $-$0.6375   & $-$0.6375  &  0.4726  & $-$0.8024   \\
&{\bf DSC}   & $-$0.6368   & $-$0.6368  &  0.4696  & $-$0.8040   \\ 
\hline\hline
\end{tabular}
\end{center}
\caption {Site-normalized ground-state energy $E_{\rm tot}$, kinetic
energy $(E_t^{x}, E_t^{y})$, and potential energy $E_U$ 
in the isotropic Hubbard model with $U/t=5$ and $x=1/8$ as obtained for 
different stripe phases: vertical site-centered (VSC), diagonal 
site-centered (DSC), vertical bond-centered (VBC) and diagonal 
bond-centered (DBC). In the HF, both types of vertical stripes 
are degenerate.} 
\label{tab:16E}
\end{table}
\begin{table}[!b]
\begin{center}
\begin{tabular}{ccccccc}
\hline\hline
&            &  $E_t^x/t$  &  $E_t^y/t$  & $E_U/t$ &  $E_{\textrm{tot}}/t$  \\
\hline
&{\bf DBC}   & $-$0.8143   & $-$0.4807  &  0.4815  & $-$0.8135   \\
&{\bf DSC}   & $-$0.8098   & $-$0.4836  &  0.4793  & $-$0.8141   \\ 
&{\bf VB(S)C}& $-$0.8304   & $-$0.4776  &  0.4938  & $-$0.8142   \\   
\hline\hline
\end{tabular}
\end{center}
\caption { The same as in Table~\ref{tab:16E} but with the hopping 
anisotropy $\epsilon_t=22\%$.}
\label{tab:16Ean}
\end{table}

\begin{figure}[t!]
\begin{center}
\includegraphics[width=0.47\textwidth ]{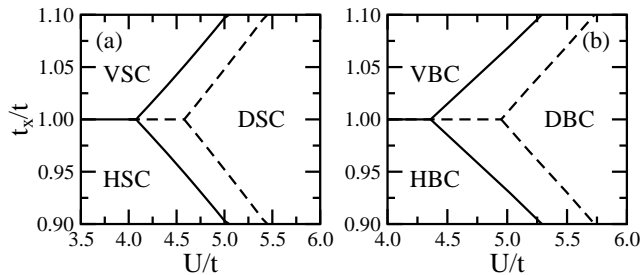}
\end{center}
\caption
{
Phase diagrams for stable:
(a) site-centered (SC), and 
(b) bond-centered (BC)
stripe structures obtained in the anisotropic Hubbard model 
on a 16$\times16$ cluster for doping $x=1/8$ (solid lines) 
and on a 12$\times12$ cluster for $x=1/6$ (dashed lines)
Parameters: $t'=0$, $V=0$.
}
\label{fig:txtyU}
\end{figure}

The effect of an increasing anisotropy illustrates the phase diagram shown 
in Fig.~\ref{fig:txtyU} determined by varying $U$ and the ratio $t_x/t_y$ 
of the nearest-neighbor hoppings in the $x$- and $y$-directions, 
while maintaining constant $t = {\textstyle \frac{1}{2}} (t_x + t_y)$. 
We observe the generic crossover from vertical to diagonal stripes with 
increasing Coulomb interaction reported in early HF studies 
\cite{Poi89,Sch89,Kat90,Inu91}. The transition from the VSC to DSC 
stripes appears in the isotropic case at $U/t\simeq 4.1$  for $x=1/8$, and 
at a higher value $U/t\simeq 4.6$ for $x=1/6$ [\textit{cf}. 
Fig.~\ref{fig:txtyU}(a)]. The corresponding phase boundary between the VBC 
and DBC stripes is shifted towards stronger Coulomb interaction and occurs 
at $U/t\simeq 4.4$ (5.0) for $x=1/8$ ($x=1/6$), respectively 
[\textit{cf}. Fig.~\ref{fig:txtyU}(b)].

The results shown in Fig.~\ref{fig:txtyU} have a simple physical 
interpretation. Stripe phases occur as a compromise between, on the one
hand, the AF interactions between magnetic ions and the local Coulomb 
interactions which favor charge localization, and the kinetic energy of 
doped holes which favors charge delocalization on the other hand. 
The kinetic energies in Table~\ref{tab:16E} show further that the vertical 
stripes are more favorable for charge dynamics. This result, which is not 
immediately obvious, has however a straightforward origin. Namely, 
the HF always leads to a large spin polarization since it is 
the only way to minimize the on-site Coulomb repulsion. Indeed, 
removal of a $\downarrow$-spin electron at site $i$ leads to relaxation of 
the $\uparrow$-spin electron energy level at this site. As a consequence, 
an alternating on-site level shift develops yielding an energetical 
motivation for the symmetry breaking and forming the AF order. 

However, the renormalization of the double-occupancy energy involves a 
strong reduction of the kinetic energy in the $\downarrow$-spin channel 
between site $i$ and its neighboring sites, as an electron incoming into 
this site encounters a high energy potential 
$U\langle n_{i\uparrow}\rangle$. Therefore, in the HF approximation we 
shall be able to identify dynamically favorable stripe patterns only by 
comparing appropriate local magnetization densities. For example, charge 
fluctuations occur more readily in the VSC stripe geometry presumably 
due to their greater overall width indicating weaker correlation effects 
(\textit{cf}. Fig.~\ref{fig:SC16}). This explains their stability at 
small $U$ where the consequent cost in potential energy $E_U$ becomes 
insignificant. By contrast, the DSC stripes are narrower having larger 
hole density along nonmagnetic DWs. Moreover, magnetization density of 
their nearest neighbor sites is markedly enhanced as compared to the 
corresponding VSC stripe magnetization, as shown in Fig.~\ref{fig:SC16} 
and in Table~\ref{tab:16SC}. The former also illustrates that the bonds 
connecting DWs with their nearest neighboring sites perpendicularly to 
the walls, have the main contribution to the kinetic energy gain, in fact 
suppressed here by larger spin polarization. Taken together, the above 
features are reflected in a more localized character of the DSC stripes, 
with a lower net double occupancy and hence a more favorable on-site 
energy $E_U$ (\textit{cf}. Table~\ref{tab:16E}). This clarifies the 
mechanism of the transition from the VSC to DSC stripes with increasing $U$.

\begin{table}[t!]
\begin{center}
\begin{tabular}{ccccccccc}
\hline\hline
         & &  $i$       & &    1   &   2   &    3  &   4   & 5     \\
\hline
         & & $\langle n^{}_{{\rm h}i}\rangle$  
         & &  0.364   & 0.234  &  0.067  & 0.014  & 0.006  \\  &{\bf VSC} &   
         & & (0.378)  &(0.234) & (0.060) &(0.013) &(0.006) \\
         & & $\langle S_i^z\rangle$    
         & &  0.000   & 0.222  &  0.348  & 0.381  & 0.384  \\   & &  
         & & (0.000)  &(0.234) & (0.357) &(0.382) &(0.384) \\
\hline
         & & $\langle n^{}_{{\rm h}i}\rangle$    
         & &  0.388   & 0.193  & 0.070   & 0.032  & 0.020   \\  &{\bf DSC} & 
         & & (0.405)  &(0.195) &(0.066) & (0.028) &(0.017)  \\
         & &  $\langle S_i^z\rangle$ 
         & &  0.000   & 0.262  & 0.352   & 0.373  & 0.380   \\  & &         
         & & (0.000)  &(0.272) &(0.360)  &(0.377) &(0.382)  \\
\hline\hline
\end{tabular}
\end{center}
\caption {Local hole $\langle n^{}_{{\rm h}i}\rangle$ and magnetization 
$\langle S_i^z\rangle$ density of the site-centered stripes shown in 
Fig.~\ref{fig:16SC}, all labeled by decreasing hole density in the 
$x$-direction. In parenthesis the values for the extended hopping model 
with $t'/t=-0.15$ are given. }
\label{tab:16SC}
\end{table}

Turning now to the analogous crossover between the BC stripes, we shall 
again compare local hole and magnetization densities on and around 
their DWs. In contrast to the SC case, a VBC stripe phase possesses 
larger hole density along DWs, as illustrated in Fig.~\ref{fig:16BC} and 
Table~\ref{tab:16BC}, suggesting that it is more localized than the DBC 
one. Nevertheless, a better renormalization of the double occupancy 
energy $E_U$ by the latter (\textit{cf}. Table~\ref{tab:16E}) follows 
from a stronger spin polarization not only of the DW atoms but also 
their nearest neighbors (\textit{cf}. Fig.~\ref{fig:16BC} and 
Table~\ref{tab:16BC}). This enhancement is directly responsible for a 
substantial reduction of the kinetic energy along bonds joining these 
atoms. Correspondingly, it accounts for a crossover from the DBC to VBC 
stripes in the small $U$ regime when the larger kinetic energy gain 
becomes crucial.

\begin{table}[b!]
\begin{center}
\begin{tabular}{cccccccc}
\hline\hline
         & &  $i$       & &    1   &   2   &    3  &   4        \\
\hline
& {\bf VBC}  & $\langle n^{}_{{\rm h}i}\rangle$  
          & &  0.326   & 0.136  &  0.030  & 0.007   \\   
          & & $\langle S_i^z\rangle$    
          & &  0.118   & 0.301  &  0.371  & 0.384   \\   
\hline
         & & $\langle n^{}_{{\rm h}i}\rangle$    
         & &  0.314   & 0.115  & 0.047   & 0.023    \\  &{\bf DBC} & 
         & & (0.323)  &(0.110) &(0.046) & (0.021)   \\
         & &  $\langle S_i^z\rangle$ 
         & &  0.145   & 0.322  & 0.365   & 0.378   \\  & &         
         & & (0.155)  &(0.333) &(0.368)  &(0.380)   \\
\hline\hline
\end{tabular}
\end{center}
\caption {The same as in Table~\ref{tab:16SC} but for the bond-centered 
stripes. VBC stripe is unstable in the extended hopping model with 
$t'/t=-0.15$.}
\label{tab:16BC}
\end{table}

We would like to emphasize that the above transition between different 
types of stripe phases is not an artefact of the HF and occurs 
also between filled stripes obtained within more realistic approaches 
including local electron correlations. Indeed, slave-boson studies 
of the Hubbard model at the doping $x=1/9$ have
established that the transition from the filled VSC to DSC stripe phase 
appears at the value $U/t\simeq 5.7$, being much higher than that 
predicted by the HF, which yields $U/t\simeq 3.8$ \cite{Sei98}. 
In this method, enhanced stability of the VSC stripes follows from an 
additional variational parameter per each site $d_i$, reducing the 
on-site energy without a strong suppression of the kinetic energy.  
Remarkably, the total energy difference between the vertical SC and BC 
stripes at both doping levels is comparable to the accuracy of the 
present calculation. Such degeneracy was also reported in the HF studies 
of the charge-transfer model \cite{Miz97}. However, when electron
correlations are explicitly included the BC stripes are more stable at
and above $x=1/8$ doping \cite{Fle01,Wre04}.

\subsection{\label{sec:3c} Effect of the next-neighbor hopping $t'$}

We now turn to the effect of a next-neighbor hopping $t'$ on the relative 
stability of the stripes. There are numerous experimental and theoretical 
results which support the presence of finite $t'$ in the cuprates. For 
example, recent slave-boson studies have revealed that the phenomena of 
the half-filled vertical stripes in LSCO requires a finite next-neighbor 
hopping $t'/t\simeq -0.2$ \cite{Sei04}. 

Let us pause now for a moment to clarify the influence of $t'$ on the 
DOS as well as on the FS using the electronic band which follows from 
a simple tight-binding model,
\begin{equation}
E({\bf k}) = -2t(\cos k_x+\cos k_y) - 4t'\cos k_x\cos k_y.
\label{eq:ek}
\end{equation}
By the reduction from the CuO$_2$ multiband model to an effective
single-band model it has been found that $t>0$ and $t'<0$ for hole doped 
system, and $t<0$ and $t'>0$ in electron doped system \cite{Fei96}. 
Although an accidental cancellation of the various contributions results 
in almost perfect electron-hole symmetry of the nearest-neighbor hopping 
$t$, the next-neighbor hopping $t'$ asymmetry appears owing to the fact 
that the dominant contribution to the latter comes from a direct O-O 
hopping $t_{pp}$ in the case of a hole hopping. On the contrary, an 
electron hopping follows from a third order 
Cu$\rightarrow$O$\rightarrow$O$\rightarrow$Cu process, being therefore 
dominated by the Cu-O hopping element $t_{pd}$.

\begin{figure}[t!]
\begin{center}
\includegraphics[width=0.47\textwidth ]{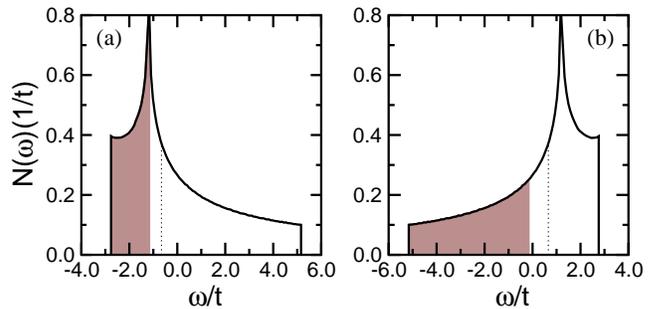}
\end{center}
\caption
{
Effect of the next-neighbor hopping $t'/t=-0.3$ on the noninteracting 2D 
DOS at the doping $x=1/4$: 
(a) hole doping ($t=1$); (b) electron doping ($t=-1$). 
Dotted line shows the Fermi energy in the undoped case, whereas the gray 
area shows the states occupied by either electrons (a) or holes (b).
}
\label{fig:DOS}
\end{figure}

In the noninteracting limit the role of $t'$ is to shift the van Hove 
singularity away from the middle of the band, either to higher or to 
lower energy depending on its sign \cite{Fle97}. Fig.~\ref{fig:DOS} shows 
the tight-binding DOS, centered at $\omega=0$ with the 
condition $\int N(\omega)\omega d\omega=0$, and the occupied states at 
the doping $x=1/4$. In the hole-doped case, with the vacuum as the zero 
electron state, the van Hove singularity lies in the lower part of the 
band. Conversely, in the case of electron doping, 
with the vacuum as the zero hole state, the van Hove singularity is 
shifted towards higher energy part of the band, unoccupied by holes.    

Apart from breaking the electron-hole symmetry, the extra parameter $t'$ 
modifies the shape of the FS of the free electrons and indeed
it becomes more consistent with the FS topology seen by ARPES 
\cite{Kin93,Arm02,Ino02}. In the electron-doped system NCCO, the 
low-energy spectral weight at the doping $x=0.04$ is concentrated in  
small electron pockets around the ($\pm\pi,0$) and ($0,\pm\pi$) points. 
Upon increasing doping, one observes both the modification of the hole 
pockets and the emergence of new low-lying spectral weights around 
($\pm\pi/2,\pm\pi/2$). Finally, at $x=0.15$ the FS pieces evolve into a 
large holelike curve centered at $M=(\pi,\pi)$. In contrast, it has been
observed that in the lightly doped regime ($x=0.03$) doped holes 
in LSCO enter into the hole pockets around ($\pm\pi/2,\pm\pi/2$) points 
\cite{Yos03}, implying that the FS is holelike and centered at the $M$ 
point. However, in the heavy overdoped regime $x=0.3$ it converts into 
the electronlike FS around the $\Gamma=(0,0)$ point.

\begin{figure}[t!]
\begin{center}
\includegraphics[width=0.47\textwidth ]{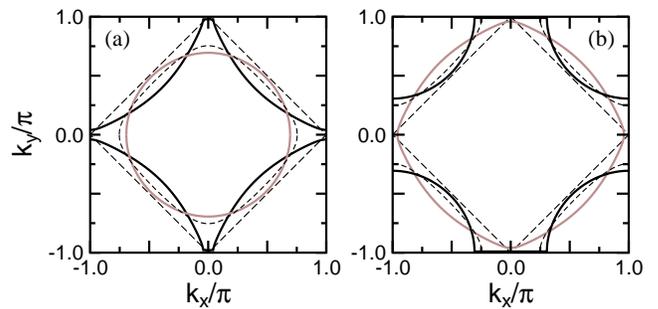}
\end{center}
\caption
{ FS obtained in the tight binding model at the doping $x=1/4$: 
(a) hole doping with $t=1$ and: $t'=-0.3$ (black solid line), $t'=0.3$ 
(gray solid line), and $t'=0$ (dashed line); 
(b) electron doping with $t=-1$ and: 
$t'=0.3$ (black solid line), $t'=-0.3$ (gray solid line), and 
$t'=0$ (dashed line). 
The long-dashed line in both panels corresponds to the undoped case 
with $t'=0$. The excessively large value of $|t'|=0.3$ as compared to 
LSCO was chosen only for more clarity of the figure. 
}
\label{fig:FSth}
\end{figure}
\begin{figure}[t!]
\begin{center}
\includegraphics[width=0.47\textwidth ]{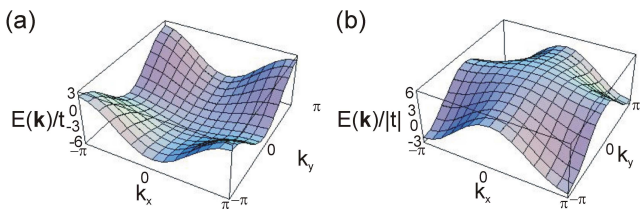}
\end{center}
\caption
{ Eigenenergy maps of the tight-binding model (\ref{eq:ek}) with 
$t'/t=-0.3$ as obtained for: 
(a) hole doping ($t=1$); (b) electron doping ($t=-1$).
}
\label{fig:ek}
\end{figure}

Fig.~\ref{fig:FSth}(a) shows that the model (\ref{eq:ek}) with $t'=0$ has 
a nested square FS at half-filling which becomes electronlike and shrinks 
around the $\Gamma$ point upon hole doping. However, negative $t'=-0.3$ 
removes the FS nesting at half filling, and the FS expands in the 
$(\pm k,0)$ and $(0,\pm k)$ directions, while contracts along the nodal 
$(k,\pm k)$ and $(\pm k,k)$ directions due to a large gradient $dE/dk$ 
along the latter. Indeed, the eigenenergy map, illustrated in 
Fig.~\ref{fig:ek}(a), has in this case a valleylike character with a 
minimum at the $\Gamma$ point. Therefore the FS turns into a holelike one 
with experimentally observed arcs [\textit{cf}. Fig.~\ref{fig:FSth}(a)]. 
In contrast, the nearest neighbor hopping $t'$ with the same sign as $t$ 
interchanges the expansion- and contraction directions 
which results in the electronlike FS.

\begin{table*}[t!]
\begin{center}
\begin{tabular}{crccrrcc}
\hline\hline
\multicolumn{1}{c} {}             &\multicolumn{1}{c}  {$t'/t$}        & 
\multicolumn{1}{c} {$E_t^x/t$}    &\multicolumn{1}{c}  {$E_t^y/t$}     & 
\multicolumn{1}{c} {$E_{t'}^{x-y}/t$}&\multicolumn{1}{c} {$E_{t'}^{x+y}/t$}  & 
\multicolumn{1}{c} {$E_U/t$}      &\multicolumn{1}{c} {$E_{\textrm{tot}}/t$}\\ 
\hline
{\bf VSC}   &  $-$0.15 & $-$0.6876   & $-$0.5886  &  0.0140  
            &  0.0140  &    0.4778   & $-$0.7704   \\   
{\bf DBC}   &  $-$0.15 & $-$0.6279   & $-$0.6279  &  0.0000 
            &  0.0183  &    0.4562   & $-$0.7813   \\ 
{\bf DSC}   &  $-$0.15 & $-$0.6275   & $-$0.6275  &  0.0000  
            &  0.0188  &    0.4533   & $-$0.7829   \\    
\hline
{\bf DBC}   &  0.15    & $-$0.6442   & $-$0.6442  &  0.0000 
            & $-$0.0282&    0.4883   & $-$0.8283   \\ 
{\bf DSC}   &  0.15    & $-$0.6437   & $-$0.6437  &  0.0000  
            & $-$0.0279&    0.4855   & $-$0.8298   \\
{\bf VB(S)C}&  0.15    & $-$0.6612   & $-$0.6372  & $-$0.0169  
            & $-$0.0169&    0.4997   & $-$0.8325   \\
\hline\hline
\end{tabular}
\end{center}
\caption {Energies per site: ground-state energy $E_{\rm tot}$, 
kinetic energy contributions for the bonds along (10) $E_t^{x}$, 
(01) $E_t^{y}$, (11) $E_{t'}^{x-y}$ and ($1\bar{1}$) $E_{t'}^{x+y}$ 
directions, as well as the potential energy $E_U$, all normalized per one
site, in the extended hopping Hubbard model with $U/t=5$ and $x=1/8$. 
VBC stripe is unstable at $t'/t=-0.15$.} 
\label{tab:16Etp}
\end{table*}

Regarding the electron doped case with $t=-1$, shown in Fig.
\ref{fig:FSth}(b), positive $t'=0.3$ (dark solid line) also leads to the 
appearance of arc segments of the FS and makes it closer to experimental 
observations. In this case, however, the minimum energy is found at 
the $M$ point, as illustrated in Fig.~\ref{fig:ek}(b). 
It should be noted in passing that this FS describes the same situation 
as the one obtained with $t=1$ and $t'=0.3$, indicated by the gray solid 
line in Fig.~\ref{fig:FSth}(a). In fact, the sign of $t$ is less important 
and turns out to be equivalent to the $(\pi,\pi)$ shift of the momentum 
without changing the corresponding eigenvalues. Consequently, in order to 
imitate the effect of hole and electron doping it is sufficient to study 
the Hamiltonian (\ref{eq:Hubb}) only below half-filling and the alternation 
between two regimes is possible by the particle-hole transform,
\begin{equation}
c^{\dag}_{i\sigma}\rightarrow (-1)^{i}c^{}_{i\sigma}, 
\label{eq:phtr}
\end{equation}
mapping the model (\ref{eq:ek}) with $t'<0$ onto the one with $t'>0$. 
Therefore, in order to avoid any further confusion concerning the signs 
of $t$ and $t'$ in Eq.~(\ref{eq:ek}), we set hereafter 
$t$ to be positive; then a negative $t'$ ($t'/t<0$) corresponds to hole 
doping, whereas a positive one ($t'/t>0$) indicates electron doping.  

The remarkable differences of the electronic structure due 
to the broken hole-electron symmetry by $t'$, result in different phase 
diagrams of LSCO and NCCO. In the former the long-range AF order is already
suppressed in the lightly doped regime $x\simeq 0.03$, while in the latter 
the antiferromagnetism is known to be quite robust at increasing electron 
doping, hence  
only {\em commensurate} spin fluctuations are observed at $x=0.15$ 
\cite{Yam03}. The robustness of the commensurate spin fluctuations in the 
electron doped regime is consistent with the ED studies of the $t$-$t'$-$J$ 
\cite{Toh94,Goo94} and $t$-$t'$-$t''$-$J$ \cite{Toh03,Toh04} models. It is
also supported by the conclusion that a negative $t'$ promotes 
incommensuration at a lower doping level than a positive one, 
reached using the QMC technique applied to the extended Hubbard model 
\cite{Duf95}. Finally, the XPS measurements in NCCO show that the 
chemical potential monotonously increases with electron doping 
\cite{Har01}, whereas its shift is suppressed in the underdoped region of 
LSCO \cite{Ino97}. These data have been nicely reproduced in 
Ref. \cite{Toh03} for both compounds, except for the low doping regime of 
LSCO where stripes are expected. All these numerical and experimental 
results indicate that doped electrons might selforganize in a different 
way than holes do --- in the latter case DWs are formed. 
Nevertheless, stable diagonal stripes with one doped electron per site in 
a DW have been obtained in the slave-boson studies of a more realistic 
extended three-band model \cite{Sad00}, so the problem is still open. 

\begin{figure}[t!]
\begin{center}
\includegraphics[width=0.47\textwidth ]{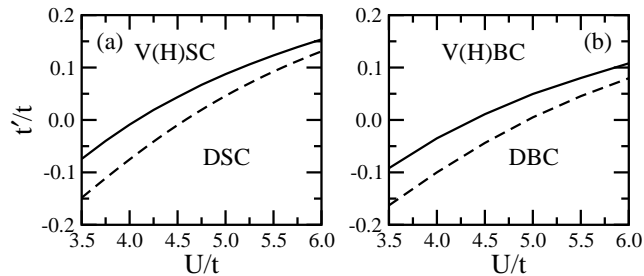}
\end{center}
\caption
{
Phase boundaries for: (a) site-centered, and (b) bond-centered stripes as 
obtained in the extended Hubbard model with the next-neighbor 
hopping $t'$ for doping $x=1/8$ (solid line) and $x=1/6$ (dashed line).  
}
\label{fig:tpU}
\end{figure}

Turning back to the competition between stripes in a doped system, 
Fig.~\ref{fig:tpU}(a) shows that negative $t'$ stabilizes the DSC stripes, 
whereas positive $t'$ favors the VSC ones, within the parameter range where 
$t'$ does not drive a stripe melting. Analogous crossover from vertical 
stripes at small $|t'|$ to more complex in shape diagonal ones at 
$t'/t=-0.1$ and $t'/t=-0.2$ has been found in other HF studies \cite{Nor02}. 
The explanation is contained in Table~\ref{tab:16Etp}: negative $t'$ gives 
a positive kinetic energy contribution, which is much more readily minimized 
by the diagonal charge configuration. Indeed, despite the solitonic 
mechanism yielding a noticeable kinetic energy loss due to the transverse 
hopping $t'/t=-0.15$, the overall kinetic energy loss in the case of DSC 
stripes along the diagonal $(11)$ and antidiagonal $(1\bar{1})$ directions 
is smaller than the corresponding one for the VSC stripe. A more careful 
analysis shows that hole propagation along the DSC stripe results in a 
contribution having the same sign as $t'$. However, it is entirely 
canceled by the ones coming from diagonal bonds of the AF domains so that 
$E_{t'}^{x-y}=0$.  

One observes further that positive $t'$ reduces the anisotropy between the
kinetic energy gains in the $x$- and $y$-directions for the VSC stripes, and 
makes their sum more favorable, while negative $t'$ has the opposite effect. 
For the DSC stripes the total kinetic energy also follows the same trend. 
The explanation of these results follows from the reinforcement of stripe 
order by a negative $t'$ (\textit{cf}. values in parenthesis in 
Table~\ref{tab:16SC}), which suppresses the hopping contributions, and its 
smearing out by  positive $t'$ where hopping is enhanced. 
These trends agree with the earlier finding within 
the DMFT that the VSC stripe phase is 
destabilized by kink fluctuations \cite{Fle01}. However, this stripe 
(dis)ordering tendency also leads to a considerably greater change in the 
Coulomb energy $E_U$, listed in Table~\ref{tab:16Etp}, for the DSC than for 
VSC stripes, which contributes significantly to the predominance of the 
former structure for negative $t'$. In fact, it follows from the increase of 
hole density within the nonmagnetic stripes and the magnetization density 
enhancement within the AF domains (\textit{cf}. Table~\ref{tab:16SC}).

Like their SC counterparts, DBC stripes are also stabilized by negative 
$t'$ resulting in a phase diagram shown in Fig.~\ref{fig:tpU}(b). In this
case, expelling holes from the AF domains enhances not only magnetization 
of their atoms but also increases magnetic moment of the hole rich DWs,
as illustrated in Table \ref{tab:16BC}. This enhancement must, however, 
strongly suppress the dominant transverse kinetic energy gain of the 
VBC stripes. Therefore, the latter are already unstable at $t'/t=-0.15$. 

It is worth noting that a finite diagonal hopping $t'$ should directly 
affect the competition between the $d$-wave pairing correlations and 
stripes. Indeed, a systematic comparison of stripe and pairing 
instabilities within the DMRG framework has shown that when the stripes 
are weakened by positive $t'$, the latter are strongly enhanced due to 
increasing pair mobility\cite{Whi99}. This effect is accompanied by a 
simultaneous enhancement of the AF correlations \cite{Toh04}. Conversely, 
negative $t'$ reinforcing a static stripe order results in the suppression 
of pair formation in the underdoped region, as found both in the DMRG 
technique and Variational Monte Carlo (VMC) \cite{Him02}. However, the 
enhanced pairing correlation, attributed to the change of the FS topology 
in LSCO, has been obtained in the optimally doped and overdoped regimes 
\cite{Shi04}.
   
\subsection{\label{sec:3d} Effect of the nearest-neighbor Coulomb 
interaction $V$}

\begin{figure}[b!]
\begin{center}
\includegraphics[width=0.47\textwidth ]{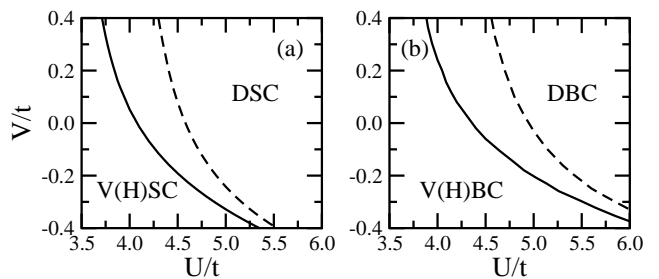}
\end{center}
\caption
{
Phase diagrams for the site-centered (a) and bond-centered (b) stripes  
obtained in the extended Hubbard model with the nearest-neighbor 
Coulomb interaction $V$ for doping $x=1/8$ (solid line) and $x=1/6$ 
(dashed line). 
}
\label{fig:VU}
\end{figure}

We now investigate the changes in the stripe stability due to either 
repulsive ($V>0$) or attractive ($V<0$) nearest-neighbor Coulomb 
interaction, which give the phase boundaries between the VSC and DSC 
stripe phases shown in Fig.~\ref{fig:VU}(a). We have found that realistic 
repulsive $V$ favors the latter. The tendency towards the DSC stripe 
formation at $V>0$ is primarily due to a large difference between charge 
densities at the atoms of the DW itself and at all their nearest-neighbor 
sites, a situation which is avoided in the case of VSC stripe phases 
(\textit{cf}. Fig.~\ref{fig:16SC}). Consequently, the former 
optimize better the repulsive  potential energy component $E_V$, as shown 
by the data reported in Table~\ref{tab:16EV}. Similarly, the fact that the 
nearest-neighbor interaction $V$ is well minimized only by inhomogeneous 
charge densities makes the DBC stripe phase more 
favorable than the VBC one, as shown in Fig~\ref{fig:VU}(b). 
While this is also the leading mechanism for both diagonal stripe 
suppression at $V<0$, the asymmetry of the curve in Fig.~\ref{fig:VU} 
arises from the fact that the lower $U$ values at the transition favor 
the higher kinetic energy contributions available for the vertical stripes.

\begin{table}[t!]
\begin{center}
\begin{tabular}{crcccrc}
\hline\hline
\multicolumn{1}{c} {}             &\multicolumn{1}{c}  {$V/t$}      &
\multicolumn{1}{c} {$E_t^x/t$}    &\multicolumn{1}{c}  {$E_t^y/t$}  & 
\multicolumn{1}{c} {$E_U/t$}      &\multicolumn{1}{c} {$E_V/t$}     &
\multicolumn{1}{c} {$E_{\textrm{tot}}/t$} \\ 
\hline
{\bf DBC}   &  $-$0.4 & $-$0.6322   & $-$0.6322   
            &  0.4626 & $-$0.6194   & $-$1.4212   \\ 
{\bf DSC}   &  $-$0.4 & $-$0.6319   & $-$0.6319  
            &  0.4602 & $-$0.6193   & $-$1.4229   \\  
{\bf VB(S)C}&  $-$0.4 & $-$0.6655   & $-$0.6083   
            &  0.4749 & $-$0.6251   & $-$1.4240   \\   
\hline
{\bf VB(S)C}&  0.4    & $-$0.6838   & $-$0.6214  
            &  0.5063 &    0.6207   & $-$0.1782   \\
{\bf DBC}   &  0.4    & $-$0.6424   & $-$0.6424   
            &  0.4829 &    0.6176   & $-$0.1843   \\ 
{\bf DSC}   &  0.4    & $-$0.6412   & $-$0.6412    
            &  0.4789 &    0.6171   & $-$0.1864   \\
\hline\hline
\end{tabular}
\end{center}
\caption {Energies per site: 
ground-state energy $E_{\rm tot}$, kinetic energy 
$(E_t^{x}, E_t^{y})$ and potential energy $(E_U, E_V)$ components 
in the extended Hubbard model with the nearest-neighbor Coulomb interaction 
$V$ for $U/t=5$ and $x=1/8$.} 
\label{tab:16EV}
\end{table}

However, it has been argued based on the results obtained using the SBA 
that an increasing repulsive interaction $V$ favors half-filled vertical 
stripes, hence the latter take over at $V/t\simeq 0.1$ in the parameter 
regime of $x=1/8$ and $U/t=10$ \cite{Sei98V}. This finding could naturally 
explain the appearance of filled diagonal stripes in the nickelates, 
provided that they were characterized by a small $V$ term, and the 
stability of the half-filled vertical ones in the Nd-codoped cuprates due 
to possibly larger value of $V$. It is also worth mentioning other HF 
\cite{Kat00} and variational \cite{Ros03} studies in which a variety of 
intriguing stripe phases, coexisting at $V/t\simeq 1.5$ with charge order, 
has been found in a broad doping region.

\subsection{\label{sec:3e} Effect of the lattice deformations}

\begin{figure}[t!]
\begin{center}
\includegraphics[width=0.47\textwidth ]{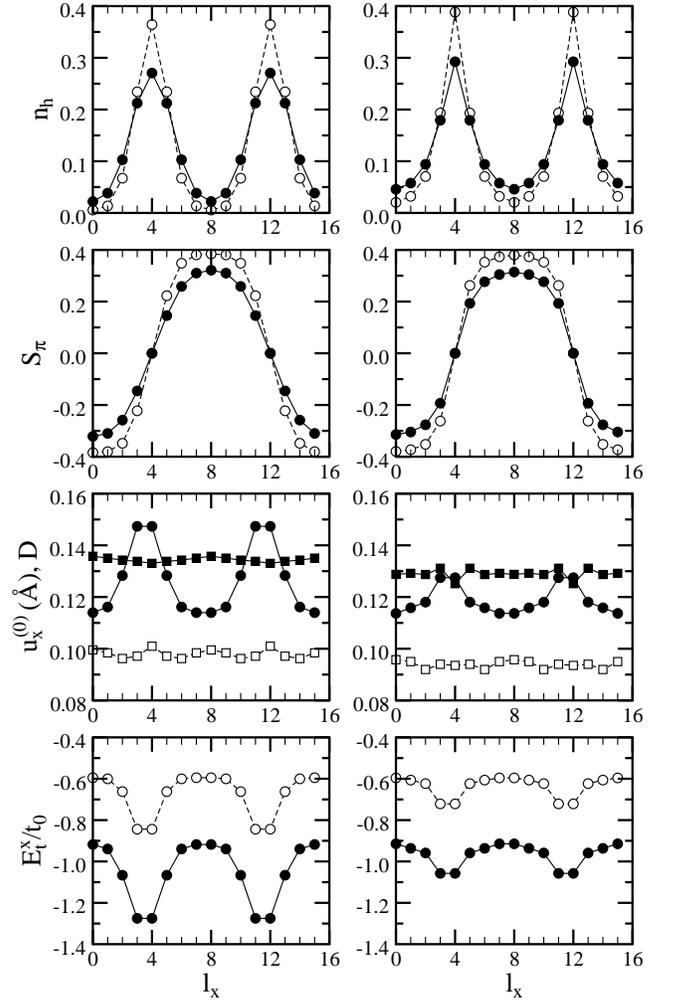}
\end{center}
\caption
{ Local hole $n^{}_{\rm h}(l_x)$ (top) and magnetization $S_{\pi}(l_x)$ 
(second row) density; fractional change of the length for the bonds to 
the right nearest-neighbor along the $x$-direction $u_{x}^{(0)}$ (circles) 
and double occupancy $D(l_x)$ (squares) (third row), as well as the 
kinetic energy $E_{t}^{x}(l_x)$ projected on the 
bonds in the $x$-direction (bottom) of
the VSC (left) and DSC (right) stripe phases, as obtained in 
the Peierls-Hubbard model (\ref{eq:hpho}) with $U/t=5$, $\lambda=0.5$ 
and $x=1/8$ (filled symbols). For comparison the results obtained with 
$\lambda=0$ are shown by open symbols. 
}
\label{fig:SC16pho}
\end{figure}

So far, we have demonstrated that a finite anisotropy of 
the transfer integral $t$ can tip the balance between vertical and 
diagonal stripes. Here we will show that such anisotropy naturally emerges 
in a doped system with DWs, described by a single-band Peierls-Hubbard
Hamiltonian,
\begin{equation}
 H= -\sum_{ij\sigma} t^{}_{ij}(u^{}_{ij})
      c^{\dag}_{i\sigma}c^{}_{j\sigma} 
       + U\sum_{i}n^{}_{i\uparrow}n^{}_{i\downarrow}
       + \tfrac{1}{2}K\sum_{\langle ij\rangle}u^2_{ij}.
\label{eq:hpho}
\end{equation}
In this model we keep only the leading term and assume a linear dependence 
of the nearest neighbor hopping element $t^{}_{ij}$ on the 
lattice displacements $u^{}_{ij}$,
\begin{equation}
  t^{}_{ij}(u^{}_{ij}) = t^{}_0(1+\alpha u^{}_{ij}).
\label{eq:tuij}
\end{equation}
Furthermore, we include the elastic energy $\propto K$ which allows to 
investigate the stability of the system with respect to a given lattice 
deformation and to determine the equilibrium configuration.  
For convenience, we parametrize the electron-lattice coupling with a single
quantity, $\lambda = {\alpha^2t_0/K}$, with the parameter values 
$K/t_0=18$\AA$^{-2}$ and $\alpha=3$\AA$^{-1}$ assumed following the 
earlier HF studies \cite{Zaa96}. As previously, we focus on the doping 
$x=1/8$ ($x=1/6$) and present the results of calculations performed on 
$16\times 16$ ($12\times 12$) clusters, respectively, with periodic 
boundary conditions. These calculations have shown that such clusters 
give the most stable filled stripe solutions for the selected doping levels. 
The model (\ref{eq:hpho}) was solved self-consistently in real space 
within the HF (\ref{eq:MF}). 
Thereby, we used an approximate saddle-point formula for the equilibrium 
relation between the actual deformation $u_{ij}$ of a given bond and 
the bond-charge density $\langle c^{\dag}_{i\sigma}c^{}_{j\sigma}\rangle$,
\begin{equation}
u_{ij}^{(0)}\simeq \frac{\alpha t^{}_0}{K}\sum_{\sigma}
\langle c^{\dag}_{i\sigma}c^{}_{j\sigma} + h.c.\rangle,
\label{eq:ph}
\end{equation}
being a consequence of the linearity assumption in Eq.~(\ref{eq:tuij}).

\begin{figure}[t!]
\begin{center}
\includegraphics[width=0.47\textwidth ]{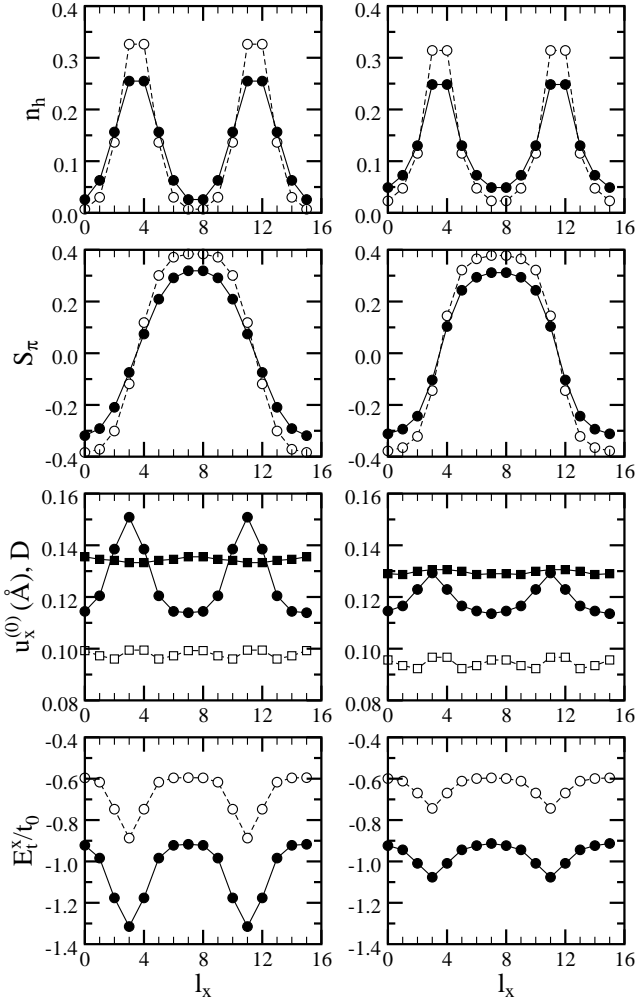}
\end{center}
\caption
{
The same as in Fig.~\ref{fig:SC16pho} but for the bond-centered stripes.
}
\label{fig:BC16pho}
\end{figure}

Quite generally, it is a widely spread out belief that inhomogeneous 
states at finite doping are very sensitive to small changes of $\lambda$, 
supported both by the HF \cite{Yon92,Yon93} and ED studies \cite{Dob94}.
Further, it has been shown that the electron-lattice interaction
favors  DW solutions over other possible phases, such as isolated polarons 
or bipolarons \cite{Zaa96}. Therefore, a complete discussion of the stripe 
phase stability in correlated oxides has to include the coupling to the 
lattice.   

We turn now to the most important aspect of this Section. Figs. 
\ref{fig:SC16pho} and \ref{fig:BC16pho} illustrate the effect of the 
finite electron-lattice coupling $\lambda=0.5$ on the SC and BC stripes,
respectively. Both figures give a clear demonstration that, in contrast 
to the hopping anisotropy $\epsilon_t$ (\ref{eq:et}) discussed above, 
finite $\lambda$ markedly modifies both the local hole density 
(\ref{eq:nh}) and modulated magnetization (\ref{eq:Spi}) [\textit{cf}. 
also Table \ref{tab:16SC} with \ref{tab:16SCpho} (SC stripes) and Table 
\ref{tab:16BC} with \ref{tab:16BCpho} (BC stripes)]. Basically, the 
influence of $\lambda$ resembles the effect of positive $t'$, smearing out 
the stripe order by ejecting holes from the DWs, being however much 
stronger. In fact, hole delocalization not only suppresses the 
magnetization within the AF domains, but also noticeably quenches magnetic 
moments of the BC domain walls. These trends can be understood by 
considering energy increments: the kinetic $E_t$, on-site $E_U$, 
and elastic energy $E_K$, as explained below.

\begin{table}[!t]
\begin{center}
\begin{tabular}{ccccccccc}
\hline\hline
    &          &  $i$       
    & &    1   &   2   &    3    &   4    & 5     \\
\hline
    &{\bf VSC} & $\langle n^{}_{{\rm h}i}\rangle$  
    & &  0.270 & 0.212 &  0.103  & 0.038  & 0.022  \\  
    &          & $\langle S_i^z\rangle$    
    & &  0.000 & 0.146 &  0.259  & 0.310  & 0.321  \\ 
\hline
    &{\bf DSC} & $\langle n^{}_{{\rm h}i}\rangle$    
    & &  0.292 & 0.179 &  0.094  & 0.058  & 0.046\\ 
    &          & $\langle S_i^z\rangle$ 
    & &  0.000 & 0.193 &  0.277  & 0.305  & 0.314\\ 
\hline\hline
\end{tabular}
\end{center}
\caption {Local hole $\langle n^{}_{{\rm h}i}\rangle$ and magnetization 
$\langle S_i^z\rangle$ density at nonequivalent atoms of the SC stripe 
phases, all labeled by decreasing hole density in the $x$-direction, 
in the Peierls-Hubbard model on a $16\times 16$ cluster with $U/t=5$, 
$\lambda=0.5$ and $x=1/8$. }
\label{tab:16SCpho}
\end{table}
\begin{table}[!b]
\begin{center}
\begin{tabular}{cccccccc}
\hline\hline
    &          &  $i$       
    & &    1   &   2   &    3    &   4     \\
\hline
    &{\bf VBC} & $\langle n^{}_{{\rm h}i}\rangle$  
    & &  0.255 & 0.156 &  0.063  & 0.026   \\  
    &          & $\langle S_i^z\rangle$    
    & &  0.074 & 0.209 &  0.291  & 0.319   \\ 
\hline
    &{\bf DBC} & $\langle n^{}_{{\rm h}i}\rangle$    
    & &  0.248 & 0.130 &  0.073  & 0.049   \\ 
    &          & $\langle S_i^z\rangle$ 
    & &  0.103 & 0.243 &  0.294  & 0.312   \\ 
\hline\hline
\end{tabular}
\end{center}
\caption {The same as in Table~\ref{tab:16SCpho} but for the BC 
stripe phases. }
\label{tab:16BCpho}
\end{table}

One should realize that a system described by the Hamiltonian 
(\ref{eq:hpho}) might be unstable towards lattice deformations only if 
the covalency increase is large enough to compensate both the $E_U$ and 
$E_K$ energy cost. Without the electron-lattice coupling, a compromise 
solution is mainly reached by developing a strong magnetic order in the 
AF domains, where a possible kinetic energy gain is irrelevant, and by 
forming nonmagnetic or weakly magnetic DWs with large hole density. 
As we have already shown, transverse charge fluctuations around the DWs 
yield the leading kinetic energy contribution. However, enhanced covalency 
and mixing of the lower $\sim\epsilon_d$ and higher $\sim\epsilon_d+U$ 
energy states between a DW and the surrounding sites partly delocalize 
these states and increase double occupancy,
\begin{equation}
D(l_x)=\langle n_{(l_x,0),\uparrow}n_{(l_x,0),\downarrow}\rangle.
\label{eq:D}
\end{equation}
Indeed, in the $\lambda=0$ case, double occupancy $D(l_x)$ reaches its 
maximum at the DWs, as illustrated in Figs.~\ref{fig:SC16pho} and 
\ref{fig:BC16pho}. The only exception is the DSC stripe phase (right 
panels of Fig.~\ref{fig:SC16pho}) with the largest $D(l_x)$ in the AF 
domains. As a consequence, the latter is the most localized one with the 
smallest kinetic energy gain (\textit{cf}. Table~\ref{tab:16E}).

The situation changes when turning on the electron-lattice coupling. 
When the electrons couple to the lattice ($\lambda\neq 0$), the bonds
contract, and the saddle point values of the distortions (\ref{eq:ph}): 
$u_{ij}^{(0)}=\langle u_{ij}\rangle$ along (10) and (01) direction, 
respectively, are finite. However, a nonuniform charge distribution 
results in a different bondlength in the cluster. This is illustrated 
in Figs.~\ref{fig:SC16pho} and \ref{fig:BC16pho} showing a fractional 
change of the length for the bonds to the right nearest-neighbor along 
the $x$-direction $u_{x}^{(0)}$ (third row). Although the values of 
$u_{ij}^{(0)}$ in the AF domains are also substantial, the largest lattice 
deformations $\sim \langle c^{\dag}_{i\sigma}c^{}_{j\sigma}\rangle$ appear 
either on the bonds connecting atoms of the DWs with their nearest 
neighbors (\textit{cf}. Fig.~\ref{fig:SC16pho}), or on the bonds which join 
two atoms of the bond-centered DWs (\textit{cf}. Fig.~\ref{fig:BC16pho}). 
Accordingly, a strengthening nearest neighbor hopping (\ref{eq:tuij}) 
enables a larger kinetic energy gain on these bonds 
(\textit{cf}. bottom of Figs.~\ref{fig:SC16pho} and \ref{fig:BC16pho}).

\begin{table}[!b]
\begin{center}
\begin{tabular}{cccccc}
\hline\hline
\multicolumn{1}{c} {}             &
\multicolumn{1}{c} {$E_t^x/t$}    &\multicolumn{1}{c}  {$E_t^y/t$}  & 
\multicolumn{1}{c} {$E_U/t$}      &\multicolumn{1}{c} {$E_K/t$}     &
\multicolumn{1}{c} {$E_{\textrm{tot}}/t$} \\ 
\hline
{\bf DBC}   &  $-$0.9679   & $-$0.9679   
            &     0.6478   &    0.2548   & $-$1.0332   \\ 
{\bf DSC}   &  $-$0.9670   & $-$0.9670  
            &     0.6450   &    0.2544   & $-$1.0346   \\  
{\bf VB(S)C}&  $-$1.0496   & $-$0.9248   
            &     0.6719   &    0.2638   & $-$1.0387   \\   
\hline\hline
\end{tabular}
\end{center}
\caption {Ground-state energy $E_{\rm tot}$ per site, kinetic energy 
$(E_t^{x}, E_t^{y})$ and potential energy $(E_U, E_K)$ components, as 
obtained in the Peierls-Hubbard model. Parameters: $U/t=5$, $\lambda=0.5$, 
and $x=1/8$.} 
\label{tab:16Eal}
\end{table}

As expected, the increasing covalency is accompanied by partial quenching 
of magnetic moments. In order to appreciate this tendency, let us 
consider a site in the AF domain with larger density of $\uparrow$-spin 
electrons (at $A$ sublattice). Once the magnetization is reduced, the 
corresponding $\uparrow$-spin energy level which belongs to the lower 
Hubbard band is pushed upwards, and the $\downarrow$-spin of the upper 
Hubbard band goes down. As a result, the locally raised $\uparrow$-spin 
state becomes stronger mixed with $\downarrow$-spin states at the 
surrounding sites of $B$ sublattice, and simultaneously bond-charge 
density increases. At the same time, electrons, jumping forth and back 
between the central site with the $\uparrow$-spin polarization and its 
nearest neighbors with the $\downarrow$-spin one, enhance considerably 
double occupancy $D(l_x)$, as shown in Figs.~\ref{fig:SC16pho} and 
\ref{fig:BC16pho}. This weakens the stripe order and results in a more 
uniform distribution of $D(l_x)$.
Of course, the increase of the elastic energy and concomitant enhancement 
of the on-site energy, both owing to finite bond contractions (\ref{eq:ph}), 
is compensated by the kinetic energy gain and the total energy is lowered
(\textit{cf}. Tables \ref{tab:16E} and \ref{tab:16Eal}).

\begin{figure}[t!]
\begin{center}
\includegraphics[width=0.47\textwidth ]{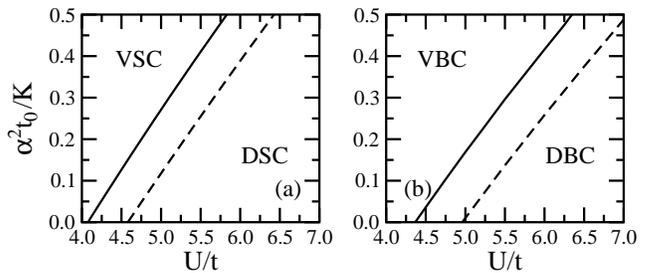}
\end{center}
\caption
{
Phase diagrams for site-centered (a) and bond-centered (b) stripe 
structures as calculated from the Peierls-Hubbard model 
for doping $x=1/8$ (solid line) and $x=1/6$ (dashed line). 
}
\label{fig:alU}
\end{figure}

We close this Section with the phase diagrams shown in Fig.~\ref{fig:alU}. 
They were obtained by varying $U$ and the coefficient $\alpha$, while 
maintaining constant $K/t_0=18$\AA$^{-2}$. The increased stability of 
vertical stripes follows from the relative stronger enhancement of the 
local hopping elements (\ref{eq:tuij}) (and consequently larger gain of 
the kinetic energy), especially on the bonds in the direction 
perpendicular to the DWs itself.

\section{\label{sec:4} Summary}

In summary, we have shown that a competition between magnetic energy of
interacting almost localized electrons and the kinetic energy of holes 
created by doping leads to the formation of new type of coexisting 
charge and spin order --- the stripe phases. We have shown that vertical 
(horizontal) and diagonal stripes dominate the behavior of the charge 
structures formed by doping the antiferromagnet away from half filling,
using the solutions obtained for the Hubbard model within the HF 
approximation in the physically interesting regime of the Coulomb 
interaction. The detailed charge distribution and the type of stripe 
order depend on the ratio $U/t$, on the value of the next-neighbor hopping 
$t'$, and on the nearest-neighbor Coulomb interaction $V$. 
We have also shown that a strong electron-lattice coupling might be 
responsible for the appearance of the vertical stripes observed in the 
superconducting cuprates at $x=1/8$. 

Altogether, although some experimentally observed trends could be 
reproduced already in the HF approach, the presented results indicate that 
strong electron correlations play a crucial role in the stripe phases and 
have to be included for a more quantitative analysis. Further progress 
both in the experiment and in the theory is necessary to establish the 
possible role of stripes in the phenomenon of high temperature 
superconductivity.

\section{\label{sec:5} Acknowledgments}
M. Raczkowski was supported by a Marie Curie fellowship of the
European Community program under number HPMT2000-141.
This work was supported by the the Polish
Ministry of Scientific Research and Information Technology, 
Project No. 1~P03B~068~26, and by the Minist\`ere
Fran\c{c}ais des Affaires Etrang\`eres under POLONIUM 09294VH.

\end{document}